\documentclass[11pt,preprint,authoryear,sort&compress]{elsarticle}

\usepackage{amsfonts,mathrsfs, amssymb, amsthm}
\usepackage{amsmath}
\usepackage{anysize}
\usepackage{array}
\usepackage[english]{babel}
\usepackage{booktabs}
\usepackage{calc}
\usepackage{caption}
\usepackage{changepage}
\usepackage{color}
\usepackage{comment}
\usepackage{etoolbox}
\usepackage{fancyhdr}
\usepackage{float} 
\usepackage{harpoon}
\usepackage[utf8]{inputenc}
\usepackage{lastpage}
\usepackage{longtable}
\usepackage{multirow}
\usepackage{natbib}
\usepackage{nicefrac}
\usepackage{paralist}
\usepackage{pbox}
\usepackage{pdflscape}
\usepackage{rotating}
\usepackage{setspace}
\usepackage{tabularx}
\usepackage{threeparttable}
\usepackage{todonotes}
\usepackage{url}
\usepackage{verbatim}
\usepackage{hyperref}

\addto\captionsenglish{}

\allowdisplaybreaks

\defcitealias{bmwi.2018}{BMWi, 2018}
\defcitealias{bmwi.2017}{BMWi, 2017}
\defcitealias{Bnetza.2017}{BNetzA and BKartA, 2017}
\defcitealias{estorage.2015}{eSTORAGE, 2015}
\defcitealias{Fraunhofer.2014}{Fraunhofer, 2014}
\defcitealias{OPSD.2017}{OPSD, 2017}
\defcitealias{SEAI.2017}{SEAI, 2017}

\doublespace

\marginsize{3cm}{3cm}{2.5cm}{3cm}

\newcolumntype{L}[1]{>{\raggedright\let\newline\\\arraybackslash\hspace{0pt}}m{#1}}

\patchcmd{\pprintMaketitle}
  {\fi\hrule}
  {\fi\ifvoid\extrainfobox\else\unvbox\extrainfobox\par\vskip10pt\fi\hrule}
  {}{}
\newenvironment{extrainfo}
  {\global\setbox\extrainfobox=\vbox\bgroup\parindent=0pt }
  {\egroup}
\newsavebox\extrainfobox

\title{
On the economics of electrical storage for variable renewable energy sources
}
\author[a]{Alexander Zerrahn}
\author[b]{Wolf-Peter Schill}
\author[c]{Claudia Kemfert}
\address[a]{\small Corresponding author, German Instutite for Economic Research (DIW Berlin), Mohrenstr. 58, 10117 Berlin, Germany, azerrahn@diw.de  \vspace*{5pt}}
\address[b]{\small German Instutite for Economic Research (DIW Berlin), Mohrenstr. 58, 10117 Berlin, Germany, wschill@diw.de  \vspace*{5pt}}
\address[c]{\small German Instutite for Economic Research (DIW Berlin), Mohrenstr. 58, 10117 Berlin, Germany and Hertie School of Governance, Friedrichstr. 180, 10117 Berlin, Germany, ckemfert@diw.de  \vspace*{5pt}}

\begin{document}

\begin{frontmatter}

\begin{abstract}
The use of renewable energy sources is a major strategy to mitigate climate change. Yet Sinn (2017) argues that excessive electrical storage requirements limit the further expansion of variable wind and solar energy. We question, and alter, strong implicit assumptions of Sinn's approach and find that storage needs are considerably lower, up to two orders of magnitude. First, we move away from corner solutions by allowing for combinations of storage and renewable curtailment. Second, we specify a parsimonious optimization model that explicitly considers an economic efficiency perspective. We conclude that electrical storage is unlikely to limit the transition to renewable energy. \\
\end{abstract}


\begin{extrainfo}
\textit{JEL codes:}\hspace{2pt} O33, \sep Q21, \sep Q41, \sep Q42
\end{extrainfo}

\end{frontmatter}


\newpage
\setcounter{page}{1}
\section{Introduction}\label{sec:intro}
In the 2015 Paris Agreement, the world agreed on ambitious targets for reducing greenhouse gas emissions to combat climate change \citep{UN.2015}. The use of renewable energy sources is a major strategy for decarbonizing the global economy. As the potentials of hydro, biomass or geothermal energy are limited in many countries, wind power and solar photovoltaics (PV) play an increasing role. For example in Germany, often considered as a frontrunner in the use of variable renewable energy sources, the government plans to expand the share of renewable energy in gross electricity consumption to at least~$80\%$ by 2050, compared to~$36\%$ in 2017 and only around~$3\%$ in the early 1990s \citepalias{bmwi.2018}. Closing this gap requires a massive further expansion of wind and solar power.

Opposed to \textit{dispatchable} technologies like coal- or natural gas-fired power plants that can produce whenever economically attractive, electricity generation from wind and solar PV plants is \textit{variable}: it depends on exogenous weather conditions, the time of day, season, and location \citep{Edenhofer.2013, Joskow.2011}. At the same time, maintaining power system stability requires to continuously ensure that supply meets demand. The potential temporal mismatch of supply and demand raises two fundamental questions: how to deal with variable renewable energy at times when there is too much supply, and how to serve demand at times when supply is scarce \citep[cf.~also][]{brown.2018}. Evidently, electrical storage can provide a solution, for instance in the form of batteries or pumped-hydro storage plants, allowing to shift energy over time. 

In a recent analysis, \citet{Sinn.2017} argues that electrical storage requirements may become excessive and could thus impede the further expansion of variable wind and solar power in Germany. Based on historic time series of electricity demand and variable renewable energy supply, he illustrates that without storage a fully renewable electricity supply would imply not using~$61\%$ of the possible power generation from wind and solar generators. In contrast, to avoid any ``waste'' of renewable energy, storage requirements to take up renewable surplus energy\footnote{Sometimes also referred to as excess energy; \citet{Sinn.2017} uses the term ``overshooting spikes''.} quickly rise to vast numbers. Under such a strategy, current German storage installations would not allow a share of wind and solar PV in electricity demand greater than $30\%$.\footnote{In 2017, Germany had a share of wind and solar energy of around~$24\%$ in gross electricity demand \citepalias{bmwi.2018}.} And for a fully renewable electricity supply, storage requirements would be more than~$400$~times as high as the currently installed German pumped-hydro storage capacity, and also much higher than the entire European potential to build such plants \citepalias{estorage.2015}. 

These considerations deserve merit as they illustrate important properties of variable renewable energy sources. As Sinn is considered to be one of the most influential economists in Germany,\footnote{See, for instance, the 2017 ranking of the most influential economists in Germany by the large German newspaper \textit{Frankfurter Allgemeine}, \url{http://www.faz.net/aktuell/wirtschaft/f-a-z-oekonomenranking-2017-die-tabellen-15173039.html} (in German).}, his conclusions can also be expected to be widely received both in policy and academic circles. This is indicated by the fact that the article was listed among the top downloads\footnote{\url{https://www.journals.elsevier.com/european-economic-review/most-downloaded-articles}.} from \textit{European Economic Review} for several months. Downloads are strongly positively correlated with citations \citep{hamermesh.2018} and thus serve as an early indicator and proxy for academic impact. As regards public impact, Sinn's analysis was covered by several influential German newspapers and magazines\footnote{These include \textit{S\"uddeutsche Zeitung}, \textit{Die Welt}, \textit{Manager Magazin}, \textit{Handelsblatt} and \textit{Wirtschaftswoche}.}, and it achieved by the time of writing an Altmetric attention score of 40, which means the article is in the top 5\% of all research outputs scored by Altmetric.\footnote{\url{https://www.altmetric.com/details/26389715}.}

Yet the approach is based on strong implicit assumptions, two of which are particularly questionable. First, it only considers two extreme cases in which either all surplus energy is stored or none. In turn, either storage needs are excessive or an excessive share of the available renewable energy is not used. An economically efficient solution is likely to be located in between, i.e., combines some amount of storage and some renewable curtailment. Second, it does not explicitly consider an economic efficiency perspective. Sinn's approach minimizes the storage energy capacity under the constraint that renewables must satisfy a specified proportion of annual electricity demand. Yet an economically efficient solution would seek to minimize the cost to reach that specified proportion of renewables. Such a solution trades off the costs of investments into storage plants, renewables that may get curtailed at times, and other assets in the power market. 


We address, and alter, these implicit assumptions and show that their effects are significant. Both results and conclusions change substantially. When moving away from corner solutions, storage needs are up to two orders of magnitude lower in a framework otherwise identical to \citet{Sinn.2017}. Using a parsimonious optimization model with a more suitable economic objective function which leads to first-best solutions, we also find moderate storage requirements. They are even lower if we consider a future broadening of the electricity sector, that is, an additional and flexible use of renewable electricity in other sectors. 

Throughout the paper, we provide the economic intuition of what drives storage requirements and use. Variable renewable energy sources are not only \textit{variable} in supply, they are also nearly free of \textit{variable cost}. A wind or solar PV plant generates electricity whenever the wind blows or the sun shines without requiring any fuel. \textit{Curtailment} of renewable energy denotes the operation of a wind or PV plant below its actual temporary generation potential, that is, neither consuming the actually available renewable energy in the moment of generation nor storing it for later use. Analogously, also conventional power plants do not generate electricity at full capacity at all times. 

The rationale 
is the following: if electricity demand is satisfied, electrical storage can be used to take up renewable surplus energy. Yet integrating increasing amounts of such surpluses requires disproportionately growing storage capacities which are not valuable at most times \citep{Denholm.2011, Schill.2014}. Thus, a corner solution avoiding any curtailment likely leads to inefficiently high storage requirements. Instead, an efficient solution seeks to balance investments into storage, renewables that get curtailed at times, and other capacities to minimize the total cost of providing electricity.

The remainder of this paper proceeds as follows: In Section~\ref{sec:literature}, we show that Sinn's results are outliers compared to the established literature. We then replicate his findings using open data and an open software tool in Section~\ref{sec:replication}. In Section~\ref{sec:curtailment}, we extend the basic model to target solutions between the two extreme cases. In Section~\ref{sec:costmin}, we devise a parsimonious model to endogenously determine optimal storage and renewable capacities as well as renewable curtailment levels. In Section~\ref{sec:discussion}, we discuss further relevant factors and flexibility options that influence storage needs. Section~\ref{sec:conclusion} concludes that electrical storage requirements are not likely to limit the transition to renewable energy.


\section{Literature review}\label{sec:literature}
Researchers from various fields have addressed the nexus of variable renewable energy and storage. Several review papers highlight different perspectives: the economic and regulatory challenges of integrating variable renewable energy sources \citep{Perez-Arriaga.2012}, features of techno-economic models required to generate policy-relevant insights \citep{Pfenninger.2014},\footnote{As techno-economic models, we classify numerical bottom-up electricity market simulation models that explicitly incorporate relevant technical constraints.} and the role of long-term storage \citep{Blanko.2018}. A synthesis of model-based analyses suggests that electrical storage requirements for renewable energy integration are generally moderate. They may only increase substantially in scenarios approaching a fully renewable energy system \citep{Zerrahn.2017}.

For Germany, \citet{Sinn.2017} derives electrical storage capacity needs of~$2,100$~gigawatt hours (GWh) ($5,800$~GWh, $16,300$~GWh), corresponding to $0.42\%$ ($1.15\%$, $3.23\%$) of yearly electricity demand, to achieve combined shares of wind and solar power of~$50\%$ ($68\%$, $89\%$). To put these numbers into perspective, we compare Sinn's results with other studies on future electricity systems with high shares of variable renewables. 

Also for Germany, \citet{Schill.2018} determine optimal storage requirements in long-run scenarios. For~$68\%$ ($78\%$, $88\%$) variable renewables,\footnote{Corresponding to overall renewables shares of $80\%$, $90\%$, and $100\%$.} they arrive at~$55$~GWh ($159$~GWh, $436$~GWh) storage, corresponding to $0.01\%$ ($0.03\%$, $0.09\%$) of annual demand. Further results on storage needs for the German energy transition are available among policy studies \citepalias{Fraunhofer.2014}. A particularly influential study concludes that hardly any additional storage investments are necessary in Germany and Europe in the short and medium term \citep{Pape.2014}: In 2050 scenarios with European shares of variable renewables around~$40\%$ ($45\%$, $55\%$), additional storage capacity between around~$14$ and~$650$~GWh is needed (corresponding to $0.00\%$ to $0.02\%$ of annual demand), largely located outside Germany. 
A long-term climate policy study commissioned by the German environmental ministry (BMUB) also finds that around $170$~GWh of pumped-hydro storage ($0.02\%$ to $0.03\%$ of annual demand) suffice to achieve variable renewable shares between~$83\%$ and~$91\%$ \citep{Repenning.2015}.

For Europe, \citet{Scholz.2017} derive cost-minimal storage capacities corresponding to about $0.08\%$ ($0.28\%$) of yearly demand to achieve $74\%$ ($85\%$) variable renewables in a setting with equal contributions of wind and solar power. Using the same numerical model, \citet{Cebulla.2017} derive larger storage needs of around~$1\%$ of yearly load at a variable renewables share of~$80\%$ in a transmission-constrained European scenario. This number decreases to~$0.5\%$ in case of increasing transmission capacity.
In a recent long-term scenario commissioned by the German energy ministry (BMWi), storage capacities in the size of~$0.01\%$ of yearly demand are enough to achieve a pan-European renewable share of~$65\%$ \citepalias{bmwi.2017}.

For the U.S., \citet{MacDonald.2016} find that integrating up to~$55\%$~variable renewables in~2030 does not require any electrical storage. Instead, pan-U.S. geographical balancing, facilitated by transmission investments, mitigates the variability of wind and solar power. For the Pennsylvania-New Jersey-Maryland market, \citet{Budischak.2013} conclude that a large stock of electric vehicle batteries, corresponding to~$0.3\%$ of yearly overall demand, would enable pushing the renewables share to~$99.9\%$ in~$99.9\%$ of all hours. Using stationary batteries would---at higher overall cost---require even less storage capacity. \citet{Jacobson.2015} show that a fully renewable (wind and solar power contributing~$90\%$) U.S. energy system covering all end use sectors would be possible with an electrical storage capacity smaller than~$0.1\%$ of yearly electricity demand.\footnote{This number includes $13$~GWh of thermal storage coupled to concentrating solar thermal power generation. In addition, the optimal solution includes substantial heat storage capacities.} For Texas, papers with different approaches also conclude on moderate storage requirements: capacities corresponding to around~$0.02\%$ ($0.06\%$, $0.14\%$) of annual demand would suffice to integrate a combined share of wind and solar PV of~$55\%$ ($70\%$, $80\%$) with relatively low renewable curtailment \citep{Denholm.2017, deSisternes.2016,Denholm.2011}. \citet{Safaei.2015} derive optimal storage deployment of $0.10\%$ of annual demand for a $66\%$ wind power share.

\begin{figure}[htb]
\centering
\includegraphics[width=1\linewidth]{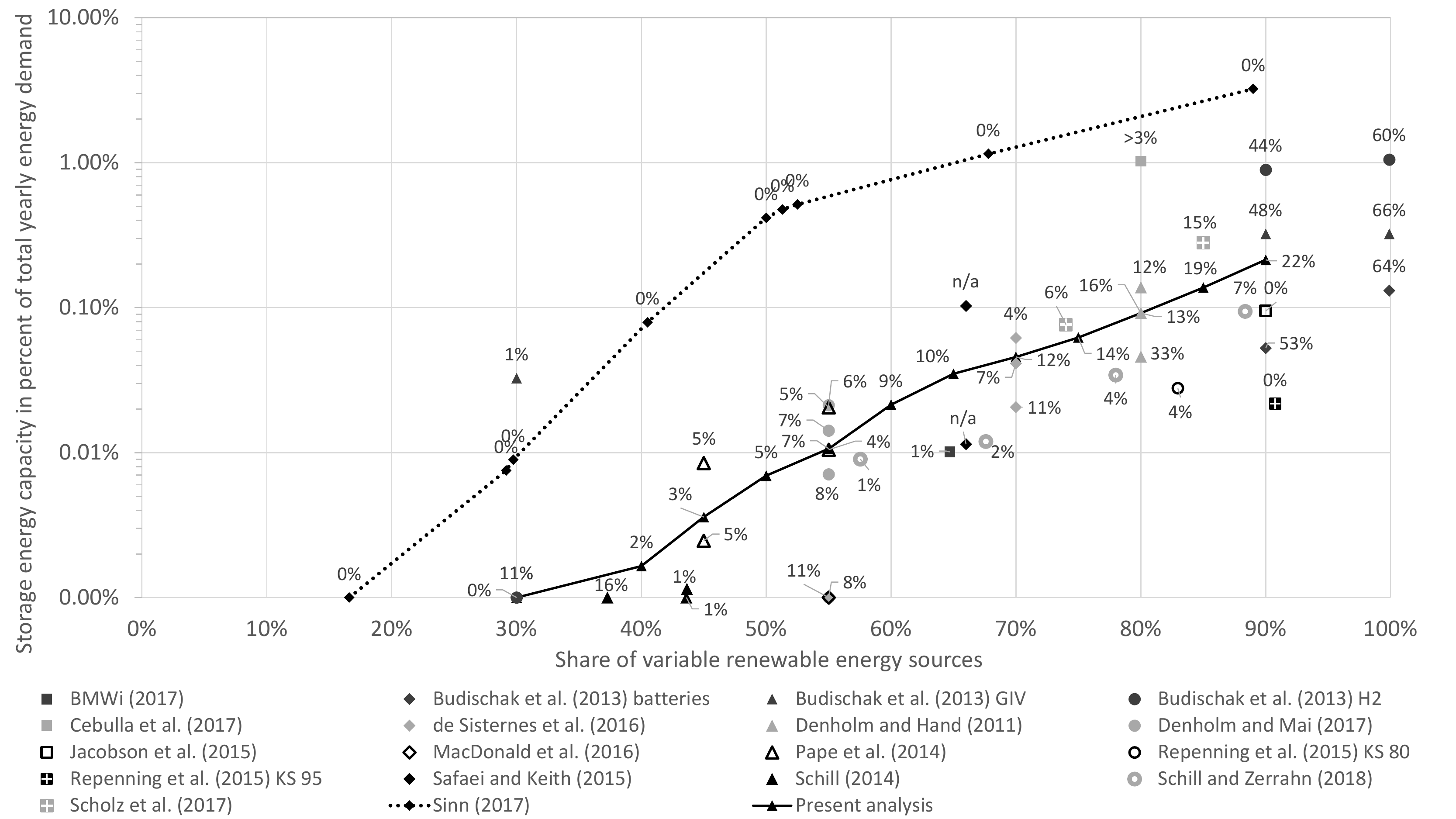}
\caption{Storage energy requirements in the literature are much lower than found by \citet{Sinn.2017}. Data labels indicate corresponding renewable curtailment in percent.}
\label{fig:survey_energy_log}
\end{figure}

Based on our review, Figure~\ref{fig:survey_energy_log} plots the shares of variable renewable energy against storage energy requirements, normalized by yearly energy demand, and contrasts them with Sinn's findings.\footnote{See \ref{app:literature_power} for more detailed information. Complementary to storage energy capacity in Figure~\ref{fig:survey_energy_log}, Figure~\ref{fig:app_survey_power} in \ref{app:literature_power} also provides additional information on storage power ratings.} It also includes the results of this paper from Section~\ref{sec:costmin}. Two findings stand out: first, storage needs disproportionately grow with higher renewables shares. Therefore, the vertical axis is provided in a logarithmic scale. This is driven by the distribution of surplus energy, which has high peaks in only a few hours of the year and is very small or zero in most other hours. Second, storage requirements found in the literature are considerably lower than those calculated by~\citet{Sinn.2017}---often by at least an order of magnitude.

\vspace*{5pt}

What drives these large differences? One important factor is renewable curtailment; the data labels in Figure~\ref{fig:survey_energy_log} provide the numbers in percent of the annually available renewable energy. Sinn focuses on corner solutions without renewable curtailment as illustrated by the outer line: the storage has to take up every potential kilowatt hour of surplus energy that could be generated by wind power and solar PV generators, which leads to strongly increasing storage requirements for growing shares of variable renewable energy sources. By contrast, the literature agrees that a complete integration of variable renewable energy is not desirable \citep[see, in particular,][]{Budischak.2013, Ueckerdt.2017, Schill.2014, Schill.2018}.

A combination of renewable capacity oversizing and some temporary curtailment substitutes storage expansion when imposing economic efficiency criteria, such as finding a technology portfolio for least-cost renewable energy supply. In equilibrium, the marginal effects of adding another unit of storage and adding another unit of renewable generation that gets curtailed at times are equal. A social planner would thus trade off storage against renewable curtailment and other options that can provide flexibility. Accordingly, renewable curtailment is not necessarily inefficient.

Beyond curtailment, further options on both the supply and demand sides can provide flexibility for variable renewable energy sources and thus substitute for electrical storage \citep{Lund.2015}. These comprise geographical balancing \citep{Fuersch.2013, Haller.2012, MacDonald.2016}, demand-side management \citep{Pape.2014, Schill.2018}, and the flexible use of renewables in other sectors such as heat or mobility \citep{Budischak.2013, Jacobson.2015}. To be concise, we largely abstract from such options in our analysis, as in Sinn's original framework. We further discuss this avenue in Section~\ref{sec:discussion}. 

Our literature review highlights two main insights: first, Sinn's findings are outliers compared to the consensus of established studies. Second, his extreme findings are driven by not considering relevant economic trade-offs concerning the provision of flexibility, in particular by neglecting potential renewable curtailment.


\section{Replication and intuition}\label{sec:replication}
We first replicate the central results of Sinn’s analysis, using a spreadsheet tool and open-source input data. Following recent discussions on good practice in the field of energy research \citep{Pfenninger.2017, Pfenninger.2018}, we provide our tools and all input parameters under a permissive open-source license in a public repository\footnote{\url{https://doi.org/10.5281/zenodo.1170554}.} 


\subsection{Focus of our analysis}\label{subsec:replication_focus}
In our replication, we focus on the central Section~6 in \citet{Sinn.2017}. Here, he derives storage requirements to integrate increasing shares of variable renewable energy from wind and solar PV in final electricity demand, ranging between~$16.6\%$ and~$89\%$. In Sections~7~and~8, he provides stylized geographical extensions of this approach; while these are illustrative, we stick to the central model and its mechanics from Section~6.\footnote{Sinn's Sections~7~and~8 illustrate storage-reducing effects of geographical balancing. These are related to smoother aggregate demand and renewable supply patterns when considering multiple countries at a time and access to flexible hydro capacities in Norway and the Alps. Yet Sinn makes a range of strong assumptions, for example, on an unchanged geographical distribution of renewables. Sinn's Section~9 provides largely qualitative reflections on sector coupling aspects. We quantitatively analyze this in our Section~\ref{subsec:costmin_p2x}.} 

In Sections~2--4, Sinn suggests transforming variable renewable supply to a perfectly constant output over all hours of a year. Such smoothing results in excessive storage needs. However, there is no economic or technical reason for this kind of smoothing. It seems to be inspired by the notion that renewable generators should mimic the characteristics of conventional power plants. In this case, additional backup capacities (referred to as ``double structures'') would be obsolete. However, it is not clear why using existing backup power plants should be less desirable than installing additional electrical storage; the approach is silent about any efficiency or optimality criteria. The lack of practical relevance is illustrated by the fact that resulting storage requirements cannot be empirically observed in countries with high variable renewables shares like Denmark, Ireland or Spain.\footnote{Especially the Irish electricity system has a high supply of wind power and at the same time only few options for intertemporal balancing. The share of wind energy in Ireland was above~$20\%$ in 2016, with wind capacities somewhat above $2,800$ megawatts \citepalias{SEAI.2017}. The only pumped-hydro storage plant had a capacity of somewhat below~$300$ megawatts and the interconnector to Great Britain a capacity of~$500$ megawatts. Compare also \citetalias{OPSD.2017}.}


\subsection{Replication using open data and an open software tool}\label{subsec:replication_replication}
We derive input data from the Open Power System Data platform, which collects and provides European electricity market data from official sources \citepalias{OPSD.2017}. Input parameters comprise hourly time series of realized German electricity demand and availability of onshore wind power and solar PV, defined as capacity factors between zero and one. Electricity demand enters the analysis as inelastic, that is, we do not fit any demand curves. This assumption follows \citet{Sinn.2017} and is also standard in much of the literature.\footnote{Assuming an inelastic short-run electricity demand appears appropriate because hourly wholesale price signals are so far not passed through to the majority of final consumers. We briefly discuss more flexible demand in Section~\ref{sec:discussion}.} Capacity factors are calculated by relating historic hourly feed-in to installed renewable generation capacity in respective hours. As in \citet{Sinn.2017}, all input data is taken from the base year 2014.

To achieve an exogenously specified share of renewable electricity, the time series of capacity factors is scaled up until renewables meet the targeted share $\delta$ of annual demand.\footnote{As inferred from \citet{Sinn.2017}, we do not add storage energy losses to overall demand throughout our analysis.} Renewables satisfy demand either contemporaneously, that is in the hour of generation, or in a following hour facilitated by storage. To this end, we impose the storage heuristic employed by Sinn: if renewable generation exceeds demand in an hour, electrical storage takes up the surplus. It is released as soon as demand net of renewable generation is positive again. As \citet{Sinn.2017}, we assume an efficiency of~$\overrightarrow{\eta}=81\%$ when storing in and~$\overleftarrow{\eta}=92.6\%$ when storing out. The remaining, non-renewable share of annual demand $1-\delta$ is supplied by some unspecified conventional technology.

The approach aims at finding the smallest possible storage size to integrate all renewable generation. It is equivalent to Sinn’s objective of scaling the storage such that it is empty in at least one hour. The hourly storage use pattern is shifted up and down until the minimum storage size is found. To avoid free lunch, we require the storage level in the first and last hour of the year to be equal. 

\begin{figure}[htbp]
\centering
\includegraphics[width=1\linewidth]{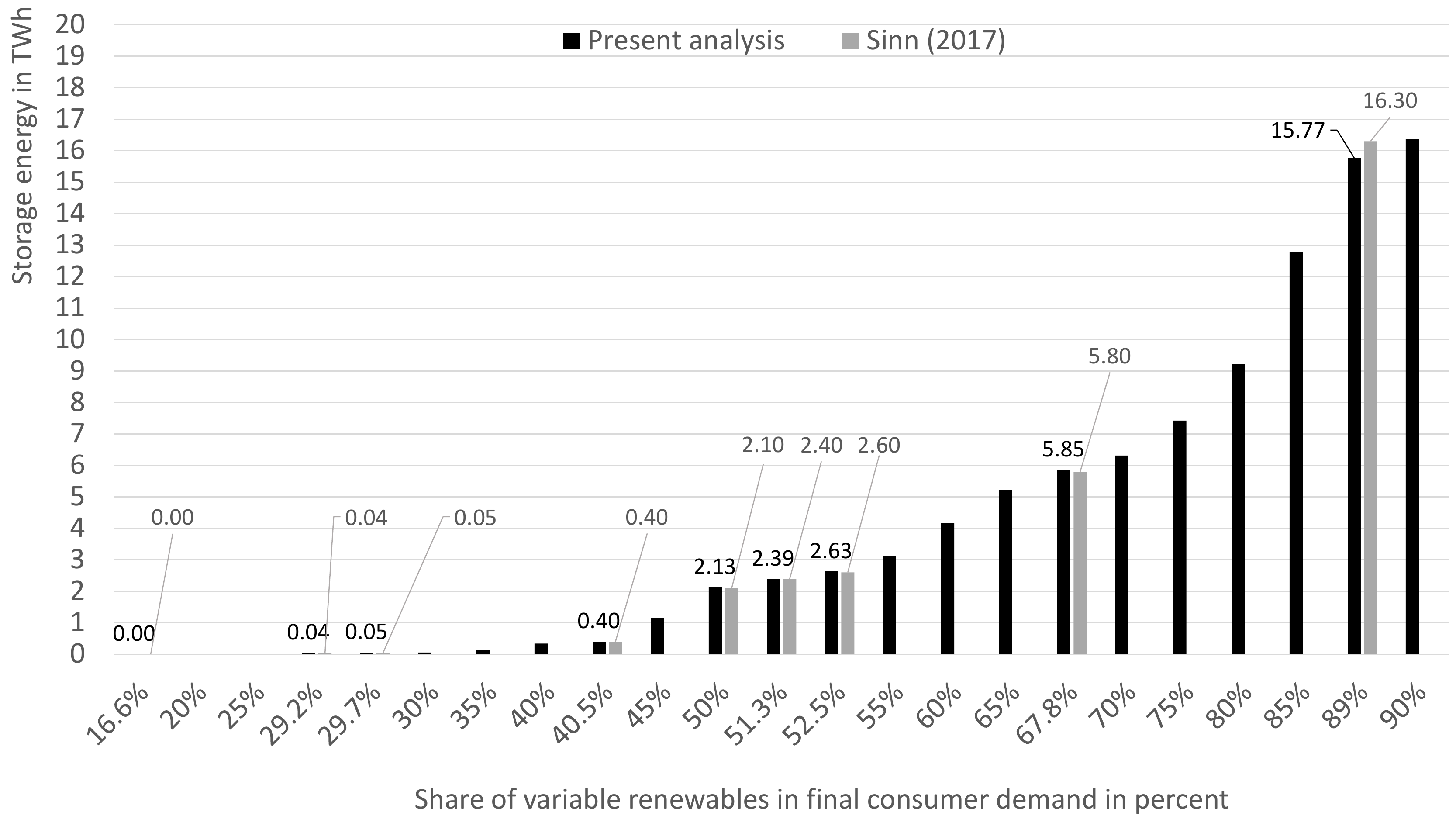}
\caption{Storage requirements rise sharply in the share of variable renewables if all renewable surplus energy must be integrated. Results replicate the findings from \citet{Sinn.2017} using open data and an open software tool.}
\label{fig:1_replication}
\end{figure}

Using open data and open software tool, we are able to replicate Sinn’s central findings.\footnote{See Table~1 in~\citet{Sinn.2017}.} Additionally, we provide results for further shares of variable renewable electricity ranging between~$20$ and~$90$~percent. Figure~\ref{fig:1_replication} shows storage requirements for varying shares of renewable electricity in final demand. Results from our calculations are given in black, Sinn’s results in gray. 

\vspace*{5pt}

Storage needs rise sharply if more renewable electricity must be integrated. While current German pumped-hydro storage installations of somewhat below $0.04$~terawatt hours (TWh) would suffice to fully integrate almost~$30\%$~renewable electricity, even moderate further increases in renewables would substantially drive up storage requirements. For~$50\%$~variable renewables, they already amount to~$2.1$~TWh, that is, they are two orders of magnitude higher. As Sinn does not provide his data and calculations open-source, we cannot trace back the small numerical differences to our findings to a specific reason; presumably, they arise due to slight differences in the input data. 


\subsection{(Non-)Robustness}\label{subsec:replication_baseyears}
We address the robustness of findings in a sensitivity analysis using different base years. Both the time series of demand and the availability of renewable energy may change substantially between years. To this end, we repeat the basic analysis using hourly time series of demand and the renewables capacity factor of the base years~2012, 2013, 2015, and 2016. 

\vspace*{5pt}
\begin{figure}[htbp]
\centering
\includegraphics[width=1\linewidth]{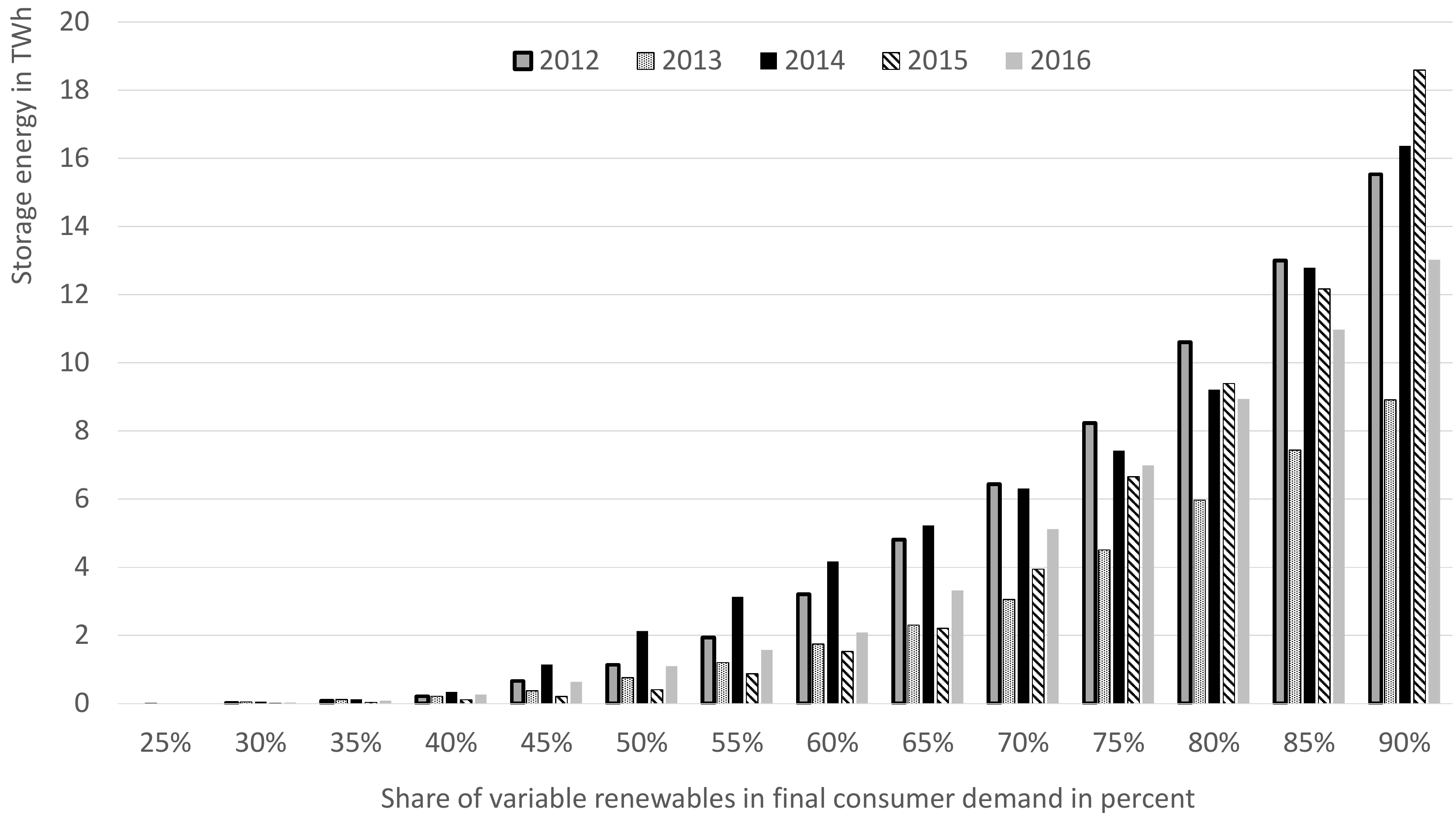}
\caption{Storage requirements are highly sensitive to the choice of the base year supplying the input data for demand and renewable availability time series.}
\label{fig:2_baseyears}
\end{figure}

Figure~\ref{fig:2_baseyears} depicts the results: storage requirements turn out to be highly sensitive to the choice of the base year. Taking data from 2014 yields the highest storage needs up to a variable renewables share of~$65\%$. In contrast, data from 2015 leads to the smallest storage sizes up to~$65\%$~renewables. For instance, comparing the~$50\%$~renewables case, storage installations are less than a fifth for 2015 data as compared to 2014 data. For a high renewable penetration beyond~$65\%$, data from the base year 2013 leads to the smallest storage. Using 2014 data generally yields comparatively large storage capacities. We explain the drivers of storage requirements below.


\subsection{Intuition: the residual load duration curve}\label{subsec:replication_rldc}
To gain intuition what drives storage requirements, we use the concept of \textit{residual load duration curves (RLDCs)}. Residual load---also referred to as net load---is defined as hourly demand minus renewable feed-in in the respective hour. It is the remaining load that conventional plants or storage installations must serve.\footnote{The expressions load and demand can be used interchangeably.} A residual load duration curve is a graphical representation of residual load of all~$8,760$~hours of a year, sorted in descending order. The positive integral of the curve corresponds to the energy that must be provided by non-renewable energy, storage generation or imports. The RLDC concept is prominent in the energy economics literature \citep[compare also][]{Ueckerdt.2017}.

\begin{figure}[htbp]
\centering
\includegraphics[width=1\linewidth]{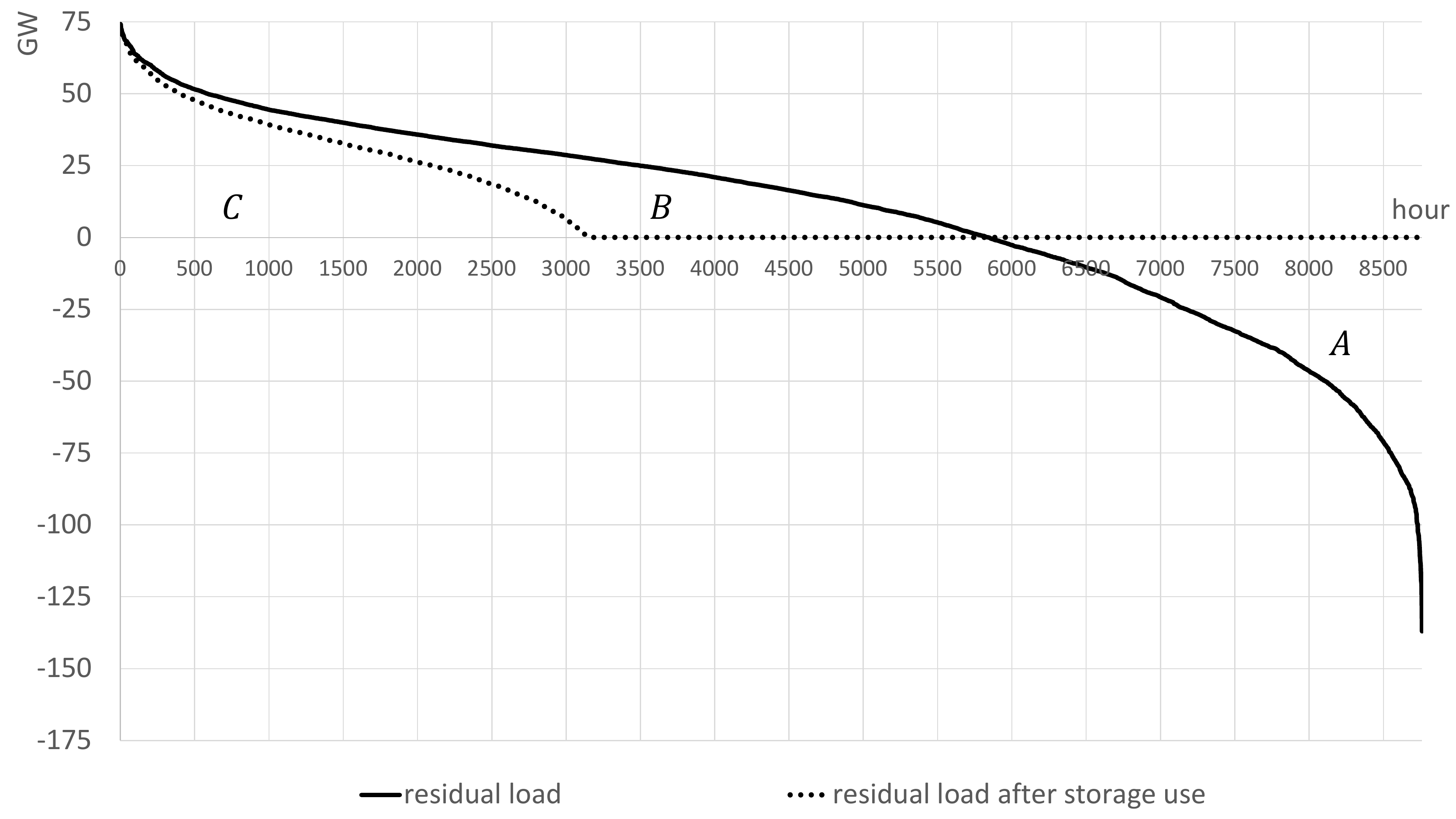}
\caption{Residual load before and after storage use for the~$80\%$~renewables case. Storage shifts surplus renewable energy (area A) to hours with positive residual load (area B).}
\label{fig:3_rldc_nocurt}
\end{figure}

Figure~\ref{fig:3_rldc_nocurt} shows RLDCs for the above analysis in case of~$80\%$~renewables, using 2014 as the base year.\footnote{We also use the base year 2014 in the remainder of this analysis, as in~\citet{Sinn.2017}.} The solid line shows residual load before storage use. To the right, below the horizontal axis, residual load is negative, see area~$A$. In these hours, there is a surplus of renewable generation. To the left, there are hours with positive residual load, areas~$B$~and~$C$. In these hours, there is relatively high demand and low renewable energy generation. The dotted line shows the RLDC after storage use.\footnote{As the residual load after storage use is also sorted in descending order, the order of hours may differ between the solid and the dotted line.} The storage takes up energy in hours of excess renewable supply and releases it in hours with excess demand. Graphically, it shifts the surplus energy represented by area~$A$ to area~$B$, which equals the size of area~$A$ reduced by the storage’s efficiency losses. Area~$C$ represents the annual residual load that must be served by other generators, for instance conventional or dispatchable renewable plants. By assumption, it corresponds to~$20\%$ of total demand in this case.\footnote{For simplicity, we do not consider electricity trade with neighboring countries. In reality, some part of the renewable surplus, area~$A$, is likely to be exported, while some part of the remaining residual load, area~$C$, is likely to be imported. The specific effects depend on trade capacities and the residual load patterns in the neighboring countries.}

\vspace*{5pt}

Importantly, nothing forces the storage to shift energy to hours with highest residual loads, that is, greatest scarcity, at the very left-hand side of the RLDC; the operational heuristic prescribes to empty the storage as soon as residual demand is positive again. With the present patterns of demand and renewable feed-in, it is unlikely that an hour with very high residual load follows closely to an hour with renewable surplus generation. The RLDC representation dissolves the temporal sequence of hours during the year.


\section{Storage requirements under renewable curtailment}\label{sec:curtailment}
The residual load duration curve illustrates the two challenges of integrating high shares of variable renewable electricity: (i) on the left-hand side of the curve, there are hours with high demand that variable renewables cannot directly supply; (ii) on the right-hand side, there are hours with renewable surplus generation. Both sides of the RLDC have an impact on storage requirements. 


\subsection{Power-oriented renewable curtailment}\label{subsec:curtailment_power}
The RLDCs suggest that curtailment of renewable surpluses may reduce storage requirements. We first devise a strategy that allows curtailment of all renewable energy surpluses beyond a defined threshold. This threshold can be interpreted as the power capacity of a storage (in~megawatt, MW); this is the energy the storage can take up per hour---as opposed to the energy capacity the storage can accommodate in total (in megawatt hours, MWh). This distinction is an essential characteristic of any electric storage technology and missing in \citet{Sinn.2017}. In case of pumped-hydro storage, the power capacity describes the power of the pumps or of the turbine to generate electricity; the energy capacity indicates the volume (in energy terms) of the storage basin.

\begin{figure}[htbp]
\centering
\includegraphics[width=1\linewidth]{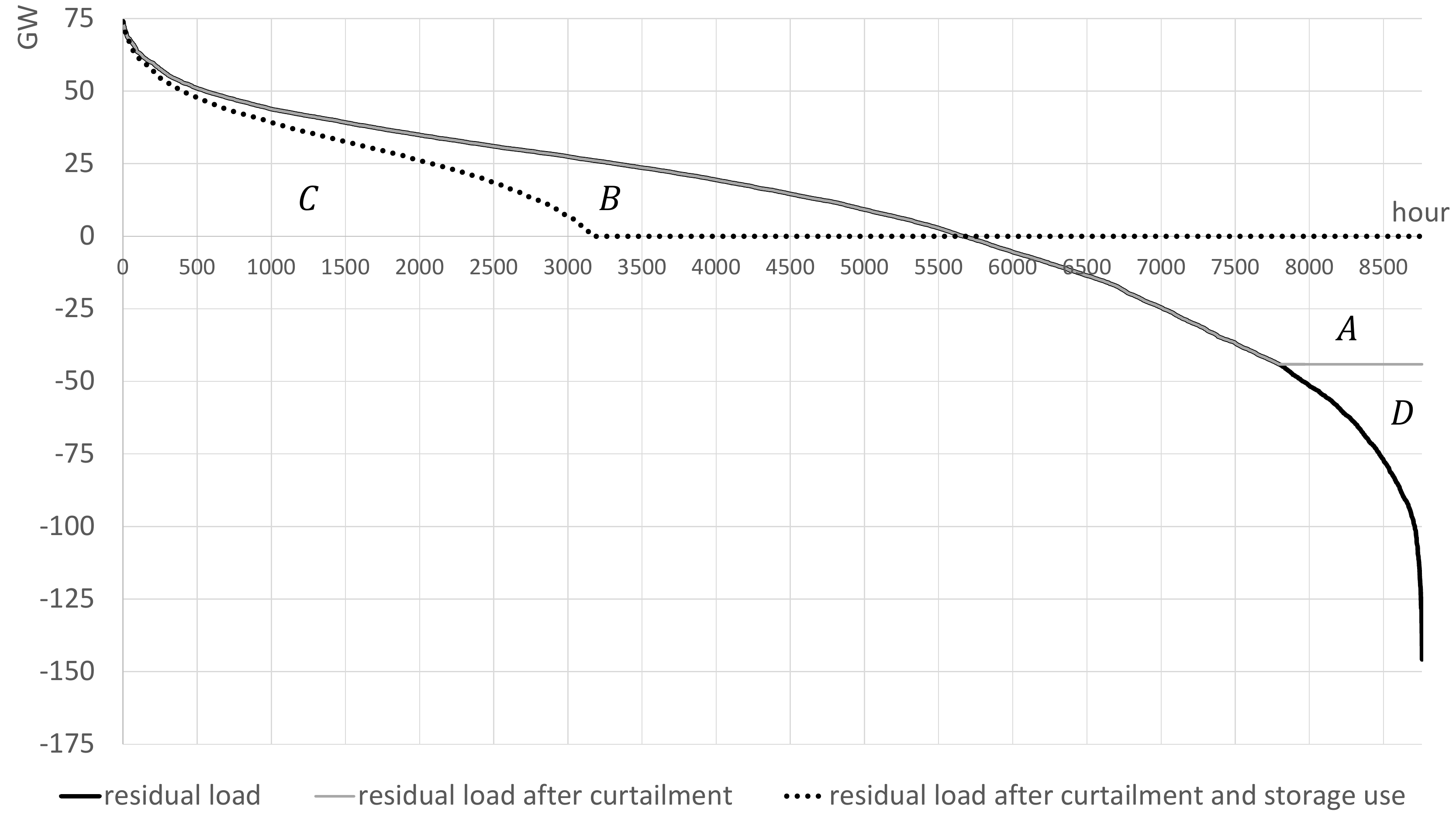}
\caption{Residual load before and after renewable curtailment as well as storage use for the~$80\%$~renewables case. Under the power-oriented renewable curtailment strategy, curtailment occurs in hours with the greatest renewable energy surplus.}
\label{fig:4_rldc_curt}
\end{figure}

We extend the basic model by a renewable curtailment threshold, equivalent to the power capacity of the storage. If hourly surplus generation is below the threshold, it is channeled into the storage; all renewable surplus energy beyond the threshold is curtailed. Otherwise, the model is identical to Section~\ref{sec:replication} and storage use remains myopic. To gain some intuition, Figure~\ref{fig:4_rldc_curt} shows RLDCs for~$80\%$~renewables in final demand and a curtailment threshold of~$44.1$~gigawatt (GW), which leads to curtailment of $5\%$ of maximum yearly renewable generation. The threshold is indicated by the horizontal solid gray part of the RLDC after renewable curtailment on the right-hand side. Area~$D$ represents all renewable surplus that is curtailed. The storage shifts the remaining surplus energy from area~$A$ to area~$B$. The remaining energy demand, area~$C$, is supplied by other means. This renewable curtailment strategy avoids storing the most excessive surplus events.
 
\vspace*{5pt}
 
\vspace*{5pt}
\begin{figure}[htbp]
\centering
\includegraphics[width=1\linewidth]{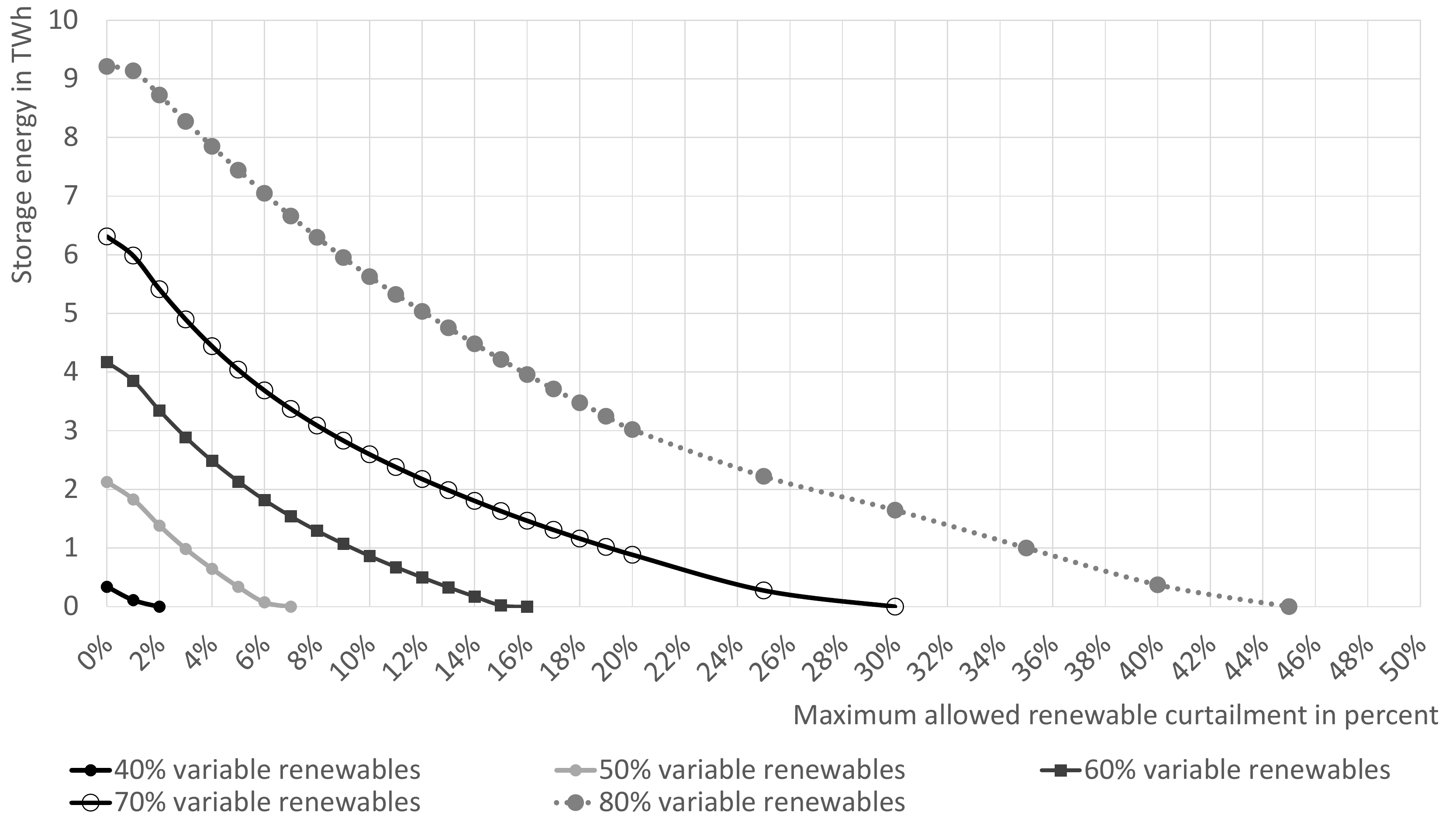}
\caption{Storage energy requirements substantially decrease when power-oriented curtailment of renewable electricity is allowed.}
\label{fig:5_less_storage}
\end{figure}

We iterate through combinations of minimum renewable requirements and maximum allowed renewable curtailment. In doing so, the spreadsheet model endogenously determines the curtailment threshold---or storage power capacity---such that no more renewable energy is curtailed than maximally allowed. Figure~\ref{fig:5_less_storage} shows the results. It is evident that increasing levels of renewable curtailment lead to lower storage requirements. The decrease is close to linear though somewhat convex. For instance, while a complete integration of~$50\%$ variable renewable electricity triggers~$2.1$~TWh storage energy capacity, allowing curtailment of $5\%$ of the annual renewable generation reduces storage needs to~$0.3$~TWh. 

Specifically, we provide solutions that lie between the two extremes ``no renewable curtailment'' and ``no storage'', which \citet{Sinn.2017} only considers. The vertical axis of Figure~\ref{fig:5_less_storage} indicates storage requirements for the corner solution if no renewable curtailment is allowed. The numbers are identical to those that we replicate from Sinn’s approach (compare Figure~\ref{fig:1_replication} and the right panel of his Table~1). The horizontal axis shows renewable curtailment levels for the corner solution if no storage is allowed. Here, we also replicate Sinn’s findings on ``efficiency losses'' that are given in the left panel of his Table~1. For instance, for~$50\%$~renewables, our model returns a renewable curtailment between~$6.0$ and~$6.5\%$, as indicated by the point where the solid gray line intersects the horizontal axis in Figure~\ref{fig:5_less_storage}. For the same case, Sinn determines an ``efficiency'' of~$93.8\%$, which corresponds to curtailment of~$6.2\%$.\footnote{For conciseness, we refrain from calculating the exact numbers here. Results on other renewables shares also replicate Sinn’s findings: in~\citet{Sinn.2017}, $40\%$~renewables correspond to~$1.7\%$~renewable curtailment if no storage is available, see left panel of his Table~1. We determine a figure between~$1.5$ and~$2\%$. For~$60\%$ ($70\%$, $80\%$) renewables, Sinn finds~$14.8\%$ ($27.1\%$, $42.6\%$) renewable curtailment and we determine a figure between~$15$ and~$16\%$ (between~$25$ and~$30\%$, between~$40$ and~$45\%$).} 

Thus, we provide a solution space combining curtailment and storage that lies between the two extreme cases. For other base years, results are qualitatively unchanged; however, they exhibit great variation concerning the level of required storage.

\begin{figure}[htbp]
\centering
\includegraphics[width=1\linewidth]{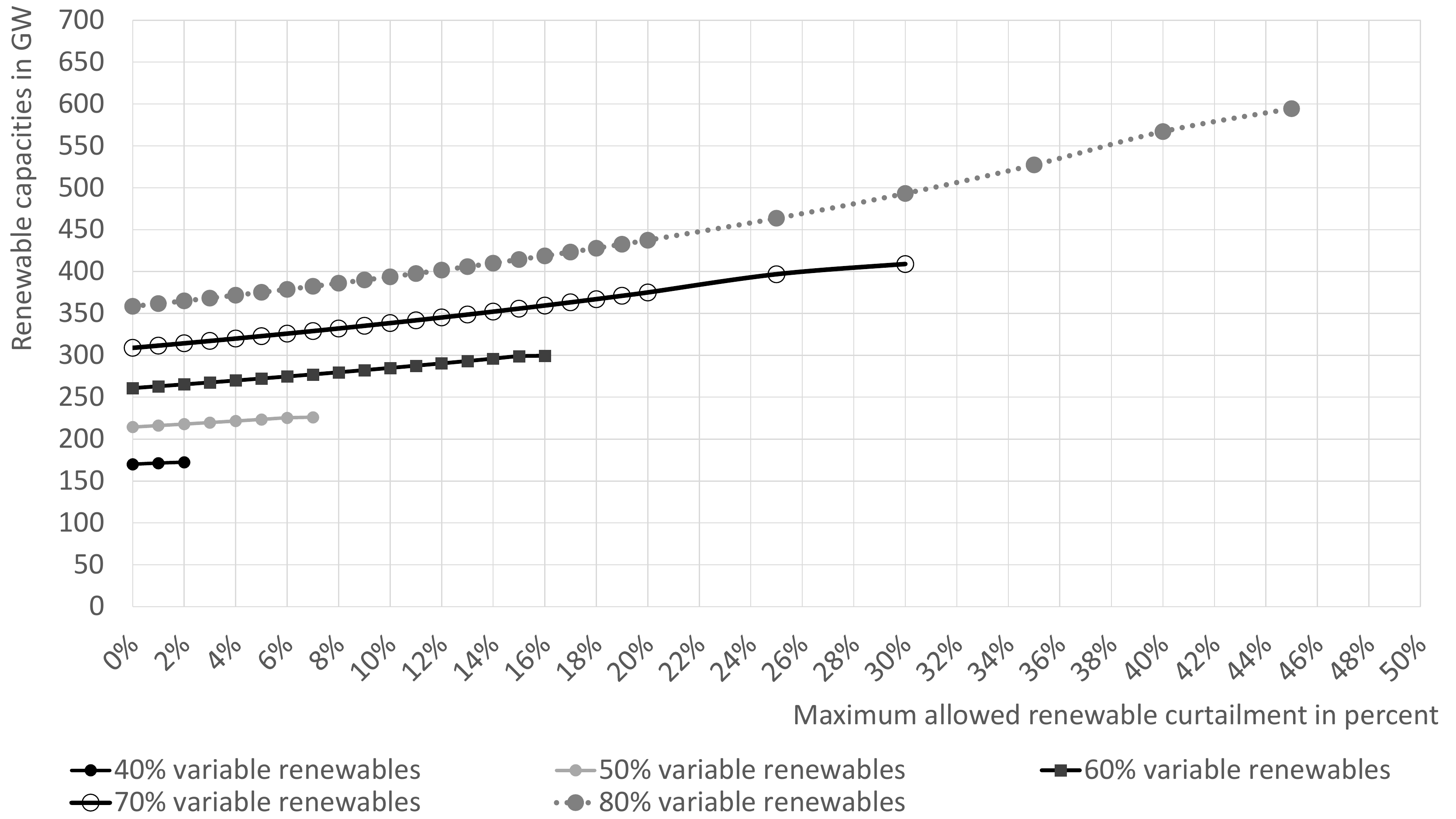}
\caption{While some renewable curtailment substantially decreases storage requirements, somewhat larger renewables capacities are necessary to achieve a specified share of renewable energy in final demand.}
\label{fig:storage_renewables}
\end{figure}

To achieve the same share of renewable energy in final demand, the required renewable capacities are necessarily higher when allowing for curtailment, that is, if some of the available energy is not used. However, this increase is moderate, as Figure~\ref{fig:storage_renewables} shows. For instance, achieving~$50\%$ renewable energy in final demand requires~$214$~GW renewables without curtailment. With~$5\%$ curtailment, the necessary renewable capacities are somewhat higher, at~$226$~GW. 

\vspace*{5pt}

Yet renewable curtailment does not increase the necessary backup capacities to supply the remaining residual electricity demand after storage. They are no larger than in the case without renewable curtailment. Inspecting the left-most part of the RLDCs in Figures~\ref{fig:3_rldc_nocurt}~and~\ref{fig:4_rldc_curt}, there is also no reason to assume so.\footnote{Backup capacities could only be smaller under the no-curtailment-regime if there was an extended number of hours with high surpluses, directly followed by hours with the highest residual loads. This case is rather unlikely; also, we cannot observe it in our data. In Section~\ref{subsec:costmin_results}, we show how storage can lower the need for backup capacities.}


\subsection{Energy-oriented renewable curtailment}\label{subsec:curtailment_energy}
While the power-oriented storage strategy---curtailing renewable surplus whenever it exceeds a defined threshold---seems plausible, it may not be optimal with respect to finding the smallest required storage energy capacity. Given historic input data, it turns out that extended periods of renewable surpluses in contiguous hours determine the maximum energy capacity of the storage, and not single periods with the highest surplus generation. For instance, storing a moderate surplus in ten consecutive hours may require more storage than storing an extreme surplus event in one hour. 

\vspace*{5pt}
\begin{figure}[htbp]
\centering
\includegraphics[width=1\linewidth]{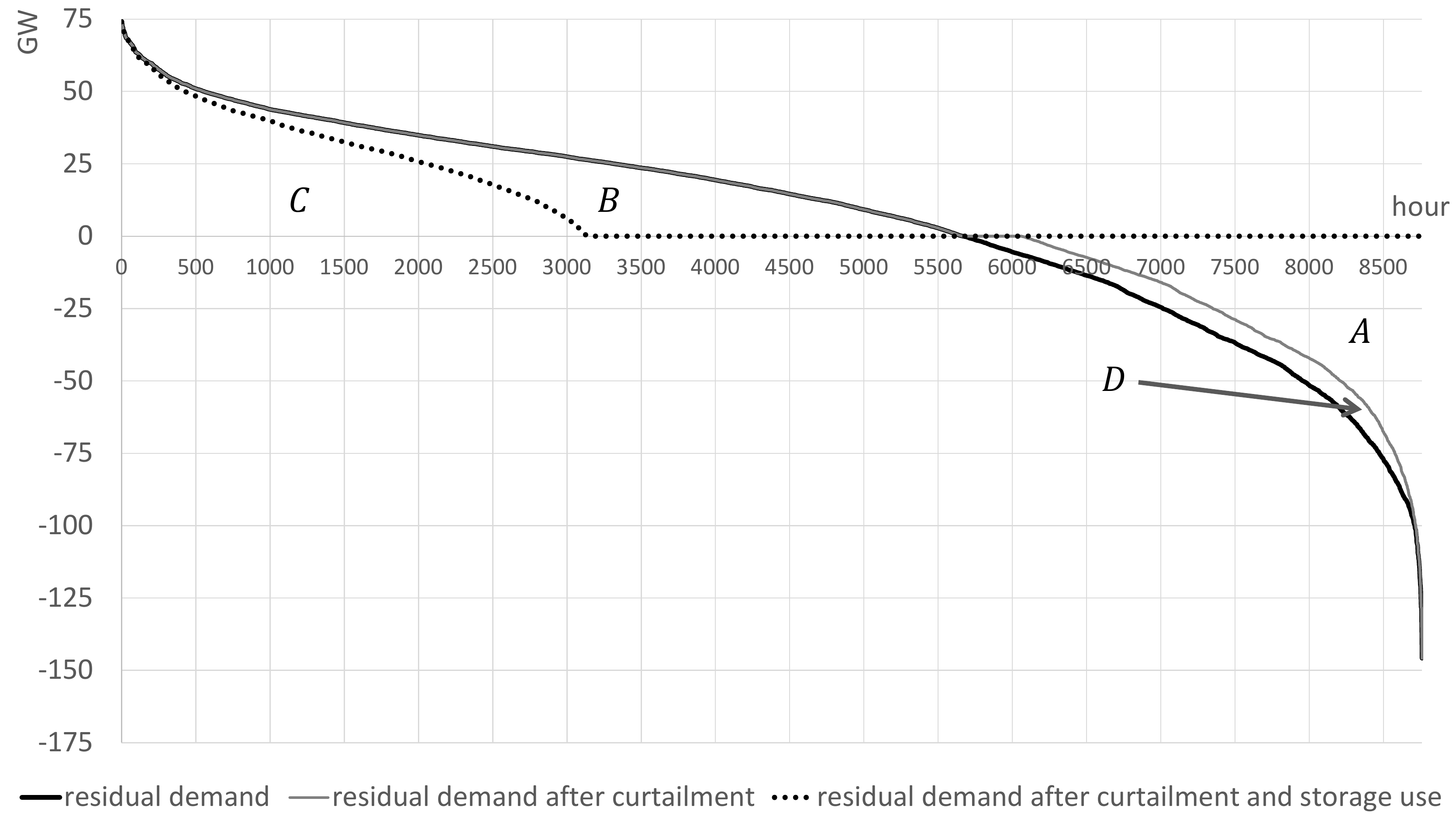}
\caption{Residual load before and after renewable curtailment as well as storage use for the~$80\%$~renewables case. Under the energy-oriented curtailment strategy, renewable curtailment occurs in hours where it triggers the greatest storage energy capacity reductions.}
\label{fig:6_rldc_energy}
\end{figure}

Therefore, we alternatively implement an energy-oriented renewable curtailment strategy. The storage operational pattern remains myopic and is identical to the above cases; however, renewable curtailment occurs if and only if the storage is fully loaded. Thus, it targets a minimum energy capacity requirement. Again, we iterate through minimum renewable requirements and maximum renewable curtailment constraints to explore the solution space and endogenously determine minimum storage capacities.

To provide some intuition, Figure~\ref{fig:6_rldc_energy} shows the resulting residual load duration curves for the case of~$80\%$~renewables and a maximum curtailment of~$5\%$ of the annual renewable energy. Curtailed energy, area~$D$, is identical to the one under the power-oriented renewable curtailment strategy, area~$D$ in Figure~\ref{fig:5_less_storage}. However, renewable curtailment is concentrated in hours in which surpluses trigger the highest storage requirements; these are not necessarily the hours with the highest surplus generation. Storage shifts the remaining surplus generation, area~$A$, to hours with positive residual load, area~$B$. Note that renewable curtailment and the storage operational pattern are still myopic and deterministic, that is, they do not require perfect foresight: surplus energy is charged into the storage as long as there are free capacities, and is curtailed otherwise. The stored energy serves residual load as soon as it is positive again.

\vspace*{5pt}
\begin{figure}[htbp]
\centering
\includegraphics[width=1\linewidth]{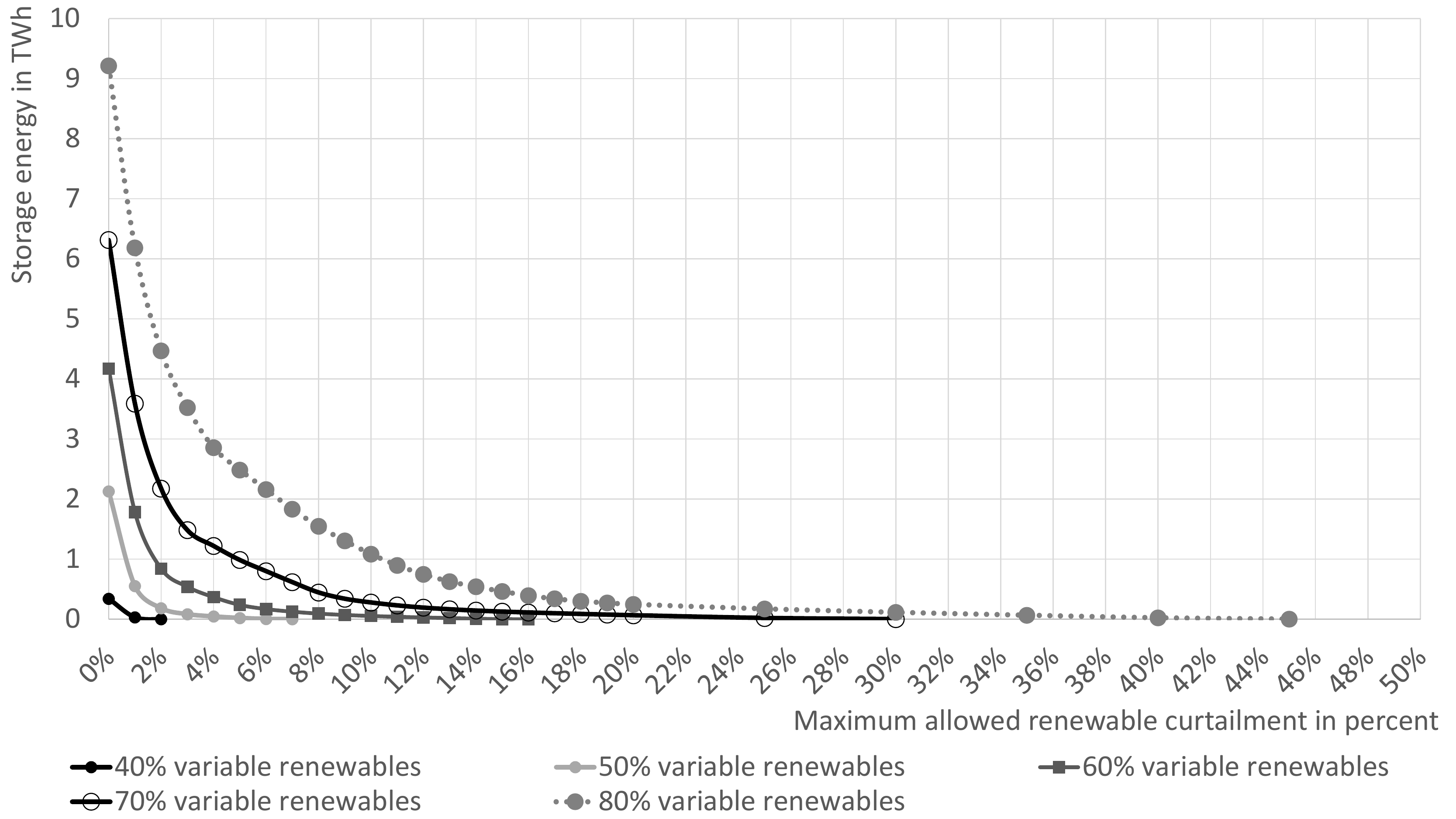}
\caption{Storage energy requirements substantially decrease if an energy-oriented renewable curtailment strategy is in place.}
\label{fig:7_less_storage_energy}
\end{figure}

Figure~\ref{fig:7_less_storage_energy} shows the resulting storage requirements for varying minimum shares of renewable energy. Overall, storage requirements decrease substantially even if only small levels of renewable curtailment are allowed. Again, the intersection points with the axes represent Sinn’s corner solutions: on the vertical axis all renewable energy must be integrated; on the horizontal axis no storage is available. Any combination of both options yields significantly lower storage needs; the decrease is much more convex than under the power-oriented renewable curtailment strategy; already a small curtailment budget triggers a large effect.\footnote{Also here, renewable curtailment does not increase backup needs. Likewise, allowing renewable curtailment goes along with slightly higher renewable capacities necessary to achieve the imposed renewables share in final demand. By construction, numbers are identical to the power-oriented curtailment strategy.} 

For instance, while a complete integration of~$50\%$~variable renewable electricity requires~$2.1$~TWh storage energy capacity, allowing for $5\%$ renewable curtailment reduces storage needs to~$0.019$~TWh, or~$19$~GWh. This is one order of magnitude lower than under the power-oriented curtailment strategy, two orders of magnitude lower than without renewable curtailment, and less than the pumped-hydro power capacity installed in Germany by 2018. Allowing~$8\%$ curtailment, $44$~GWh~of storage, slightly more than installed in Germany by 2018, would suffice to reach~$70\%$~variable renewable energy.


\section{Cost-minimal storage requirements}\label{sec:costmin}
The data-driven analysis in Section~\ref{sec:curtailment} alters an important implicit assumption of Sinn’s approach: plausible solutions lie between the two corner solutions of no renewable curtailment or no storage. However, the approach is still unlikely to result in an efficient market outcome because of the objective function used. From an economic perspective, finding least-cost solutions is relevant, that is, cost-minimal combinations of conventional and renewable generation, renewable curtailment, and electrical storage.\footnote{For the sake of conciseness and traceability, we still neglect other potential sources of flexibility, such as load shifting or dispatchable biomass. We illustrate the effects of such flexibility options in \citet{Schill.2018}.} As we also include a stylized representation of conventional generators, we explicitly address what Sinn refers to as ``double structure buffering.''

In this context, both storage energy capacity (in~MWh) and storage power capacity (in~MW) matter with respect to costs.\footnote{For simplicity, we assume identical charging and discharging capacity.} To find optimal solutions, we employ a stylized and parsimonious numerical optimization model.\footnote{The model is derived from our established and more detailed open-source model DIETER; see \citet{Zerrahn.2017} for an exposition. The model is implemented in the General Algebraic Modeling System (GAMS).} We provide the source code and all input data under a permissive open-source license in a public repository.\footnote{\url{https://doi.org/10.5281/zenodo.1170554}}

The economic optimization approach addresses both challenges of renewable energy integration. First, it delivers an efficient solution to the trade-off how much and when renewable surplus energy to curtail, and how much and when to store. This corresponds to the right-hand side of the residual load duration curve. Second, it determines efficient conventional, renewable, and storage capacities to serve demand at any point in time; this corresponds to the left-hand side of the residual load duration curve. The results of the cost minimization model can be interpreted as long-run equilibria under the assumption of perfect competition and complete information. The model thus mimics a first-best social planner approach.


\subsection{The model}\label{subsec:costmin_model}
The numerical model minimizes the total costs of satisfying electricity demand in every hour~$h$ of a year. The objective function~\eqref{eq:1_objective} sums the products of specific investment costs $\kappa^{i}$ and capacity entry $N$ of storage, differentiated by energy~$N^{se}$ and power~$N^{sp}$, renewables~$N^{r}$, and conventional capacity~$N^{c}$. Throughout the model, upper-case Roman letters indicate variables. 

For conciseness, we consider one stylized technology for variable renewables which aggregates the generation patterns of onshore wind power and solar PV,\footnote{According to the base year 2014, the synthesized renewable technology consists of~$61\%$~onshore wind power and~$39\%$~solar PV.} and two stylized conventional technologies $c \in \{base,peak\}$, parameterized to lignite and natural gas plants. Base generators incur high capacity costs and low variable costs, and vice versa for peak plants. We parameterize storage according to pumped-hydro storage, which features relatively high costs for power capacity~$\kappa^{i,sp}$ and relatively low costs for energy capacity~$\kappa^{i,se}$. By focusing on pumped-hydro storage, we follow the narrative in \citet{Sinn.2017}. Moreover, it is a mature technology and its cost structure renders it a favorable medium-term storage. This complements the variability of renewables well, especially with regard to the diurnal fluctuations of PV.\footnote{See \ref{app:battery_sensitivity} for a model sensitivity with storage parameterized to lithium-ion batteries.} Investment costs are annualized using typical lifetimes of power plants. For simplicity, we abstract from the lumpiness of investments. All variables are continuous and positive. 

Moreover, the objective function comprises operational costs for conventional plants~$\kappa^{v,c}$, consisting of fuel and other variable costs, and storage use~$\kappa^{v,s}$. These operational costs apply to each megawatt hour of conventional generation~$G^{c}_h$ and each megawatt hour of storage charge charging~$\overrightarrow{S}_h$ and discharging~$\overleftarrow{S}_h$. 

Renewable energy does not incur any variable costs. We also do not impose any costs for curtailment of renewables. As the objective function comprises the investment costs of renewable plants, it accounts for the full cost of renewable energy irrespective whether it eventually satisfies demand or is curtailed at times.\footnote{According to current German legislation, owners of wind and PV plants generally receive a subsidy payment for each megawatt hour of energy generated, also when being curtailed, to recover their investments. Analogously, the model's objective function accounts for investment costs of wind and PV plants irrespective whether generation is curtailed at times or not. In 2016, curtailment amounted to~$2.3\%$ of the renewable energy generation under the German subsidy scheme, however, entirely mandated by the electricity network operators to ease congestion \citepalias{Bnetza.2017}.} By analogy, also the use of conventional plants below capacity does not receive any penalty in the objective function.

All parameter assumptions lean on established projections for Germany for 2035.\footnote{We do not aim to derive detailed projections for a future electricity market in Germany here. Our stylized analysis only focuses on relevant drivers influencing an cost-optimal storage capacity. Therefore, we abstract from further economic and technological details.} The concrete numbers are provided in the Zenodo repository.
\begin{align}
\min Z = \kappa^{i,se}N^{se} &+ \kappa^{i,sp}N^{sp} + \sum_{c}\kappa^{i,c}N^{c} + \kappa^{i,r}N^{r} \nonumber \\
&+ \sum_{c,h}\kappa^{v,c}G^{c}_h + \sum_{h}\kappa^{v,s}\left(\overrightarrow{S}_h + \overleftarrow{S}_h\right) \label{eq:1_objective}
\end{align}

The market clearing condition~\eqref{eq:2_balance} makes sure that price-inelastic electricity demand~$d_h$ in every hour is satisfied either by renewable generation~$G^{r}_h$, conventional generation or generation from storage. 
\begin{align}
d_h = \sum_{c}G^{c}_h + G^{r}_h + \overleftarrow{S}_h \qquad \forall~h \label{eq:2_balance}
\end{align}

Constraint~\eqref{eq:maxprod} ensures that hourly generation by conventional plants does not exceed installed capacity. Hourly renewable energy supply is determined as the product of the exogenous hourly capacity factor~$\gamma_h \in \left[0,1\right]$ and the installed capacity. Both the hourly time series of electricity demand and the capacity factor enter the model as data. We take the same time series as is Sections~\ref{sec:replication} and~\ref{sec:curtailment}, that is, German data from the base year~2014. Renewable energy either satisfies demand, is charged into the storage~$\overrightarrow{S}_h$, or gets curtailed~$C^{r}_h$ \eqref{eq:3_distrib}. For convenience, we stick to the assumption that storage can only be charged with renewable energy.
\begin{subequations}
\begin{align}
G^{c}_h &\leq N^{c} \qquad \forall~h \label{eq:maxprod} \\
\gamma_hN^{r} &= G^{r}_h + \overrightarrow{S}_h + C^{r}_h \qquad \forall~h \label{eq:3_distrib}
\end{align}
\end{subequations}

The storage level~$\overline{S}_h$ in any hour equals the storage level in the previous hour~${h-1}$, plus the energy charged to storage~$\overrightarrow{S}_h$ minus the energy discharged~$\overleftarrow{S}_h$, both corrected by efficiency losses~\eqref{eq:4a_storage_move}, which are identical to the spreadsheet approach. Capacity constraints impose that the hourly energy charged or discharged does not exceed the installed pump or turbine capacity~(\ref{eq:4c_storage_maxpowerin}--\ref{eq:4d_storage_maxpowerout}) and that the storage level never exceeds the installed energy storage capacity~\eqref{eq:4b_storage_maxenergy}. Further, we require an identical storage level in the first and last period of the analysis.
\begin{subequations}
\begin{align}
\overline{S}_h &= \overline{S}_{h-1} + \overrightarrow{\eta}\overrightarrow{S}_h - \frac{\overleftarrow{S}_h}{\overleftarrow{\eta}} \qquad \forall~h \label{eq:4a_storage_move} \\
\overrightarrow{S}_h &\leq N^{sp} \qquad \forall~h \label{eq:4c_storage_maxpowerin} \\
\overleftarrow{S}_h &\leq N^{sp} \qquad \forall~h \label{eq:4d_storage_maxpowerout} \\
\overline{S}_h &\leq N^{se} \qquad \forall~h \label{eq:4b_storage_maxenergy}
\end{align}
\end{subequations}

To explore rising shares of renewable electricity, we exogenously impose a minimum share of yearly final demand to be satisfied by renewables, $\delta \in \left[0,1\right]$, for reasons of convenience imposed as maximum share of conventional energy~\eqref{eq:5_minres}. We explore minimum renewables shares between~$25\%$ and~$90\%$ in five percentage points increments.
\begin{align}
\sum_{c,h}G^{c}_h \leq \left(1-\delta\right)\sum_h d_h \label{eq:5_minres}
\end{align}

The model is a linear program and solved numerically to global optimality. The result is a cost-minimal combination of renewable, conventional, and storage capacity investments as well as their optimal hourly dispatch. Specifically, two strategies can increase the required minimum share of renewables: the use of storage to integrate surpluses, or larger renewable capacities plus curtailment. The model solves this trade-off endogenously. Note that, opposed to the myopic models in Section~\ref{sec:curtailment}, this approach requires the assumption of perfect foresight to optimally schedule the release of energy from the storage.\footnote{In the real world, proficient market forecasts are essential for storage operators. The market reality thus likely lies between the myopic and the perfect foresight cases.} 


\subsection{Intuition}\label{subsec:costmin_intuition}
\begin{figure}[htbp]
\centering
\includegraphics[width=0.95\linewidth]{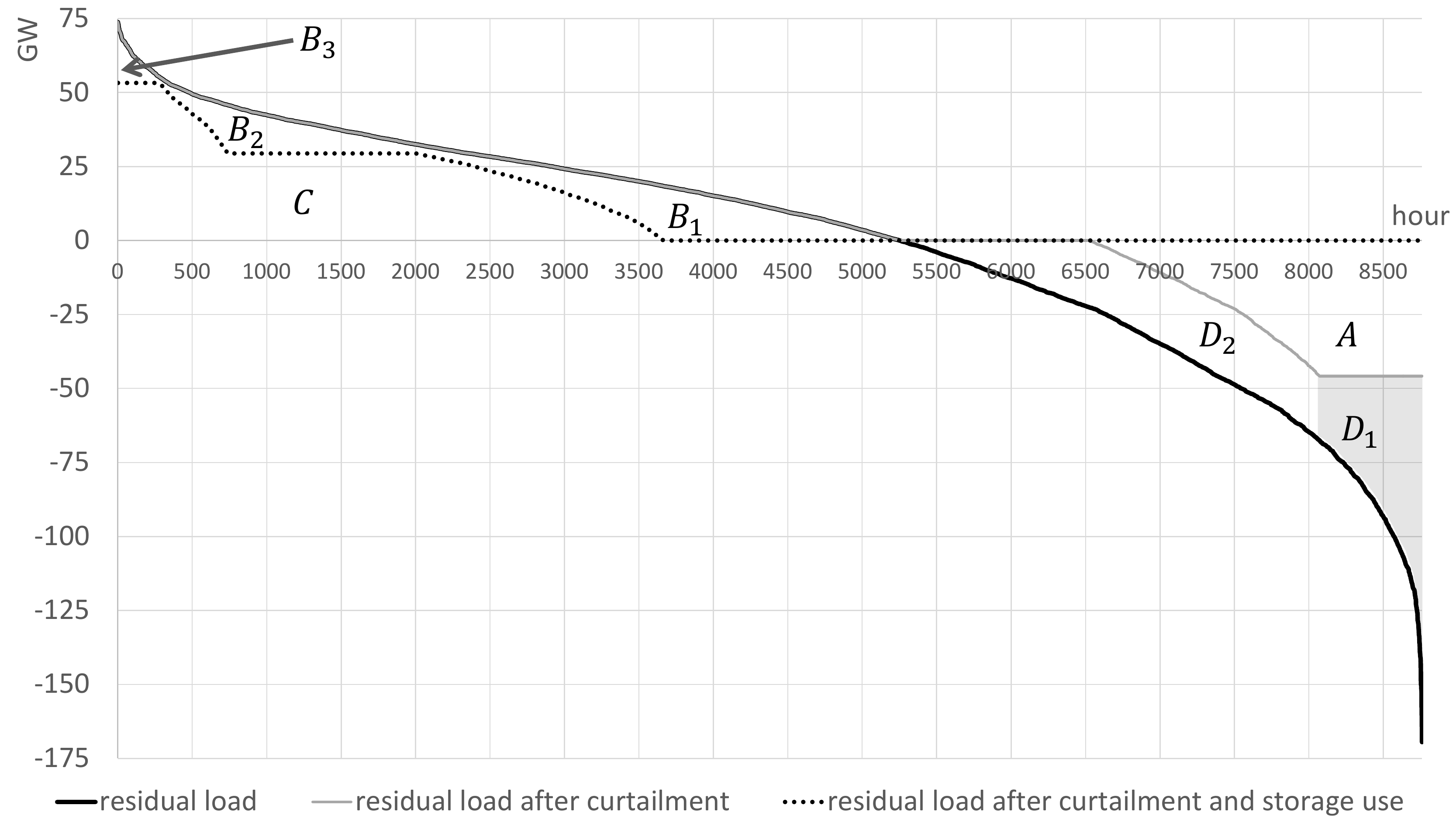}
\caption{Residual load before and after renewable curtailment as well as storage use for the~$80\%$~renewables case. In the cost-optimal solution, curtailment follows both a power- and an energy-oriented strategy. The storage shifts renewable surpluses in time to replace variable costs and investments of conventional generators.}
\label{fig:8_rldc_costmin}
\end{figure}

To again provide some intuition before discussing numerical results, Figure~\ref{fig:8_rldc_costmin} plots the resulting RLDCs for the~$80\%$ renewables case. If capacity entry is costly with respect to both storage power and storage energy, the optimal solution combines both channels analyzed in Section~\ref{sec:curtailment}. Areas~$D_1$ and~$D_2$ represent the curtailed renewable energy. The kink to the right is driven by power-oriented curtailment, compare Figure~\ref{fig:4_rldc_curt}. The renewable surplus gets curtailed beyond the threshold of~$46$~GW, as the shaded area~$D_1$ indicates, limited by the optimal storage power capacity. Curtailment of renewable surplus $D_2$ is driven by energy-oriented renewable curtailment, compare Figure~\ref{fig:6_rldc_energy}, targeted at limiting the storage energy capacity. Area~$A$ represents the stored renewable surplus energy, which the storage shifts to hours with positive residual load.
 
The optimal release of the stored renewable energy surpluses economically balances three uses: First, the energy shifted to area~$B_1$ reduces generation---and according variable costs---of the base technology, which otherwise operates whenever residual load is positive. Second, the energy shifted to area~$B_2$ reduces variable costs of the peak technology. Base capacity is~$29.4$~GW, indicated by the horizontal part of the dotted line; beyond, the peak plant with higher variable costs additionally generates electricity. Third, the energy shifted to area~$B_3$ replaces capacity entry---and respective cost---of the peak technology. If there was no storage, additional generation capacity would have to satisfy the demand exceeding~$53.5$~GW, indicated by the left-most horizontal part of the dotted line.\footnote{In fact, the storage crowds out both peak and base capacities; the optimization endogenously determines the optimal relationship. In the figure, only the dampening effect on peak capacities is clearly visible.}

\vspace*{5pt}

Thus, the cost-minimization model illustrates two economic values of electrical storage beyond avoiding renewable curtailment in a concise way. First, an arbitrage value materializes when stored renewable surplus energy replaces variable costs of other generators, in particular fuel costs. Second, a capacity value materializes when stored renewable surplus energy replaces conventional power plant capacities which otherwise would have to be provided for hours of residual load peaks. Still, only renewable energy can enter the storage by assumption. Opening up the storage for conventional energy would strengthen both economic values. In reality, no reason prohibits such operation.\footnote{Residential battery storage coupled to prosumage-oriented solar PV installations constitutes a notable exception \citep{Schill.2017b}.}


\subsection{Results}\label{subsec:costmin_results}
\begin{figure}[htbp]
\centering
\includegraphics[width=1\linewidth]{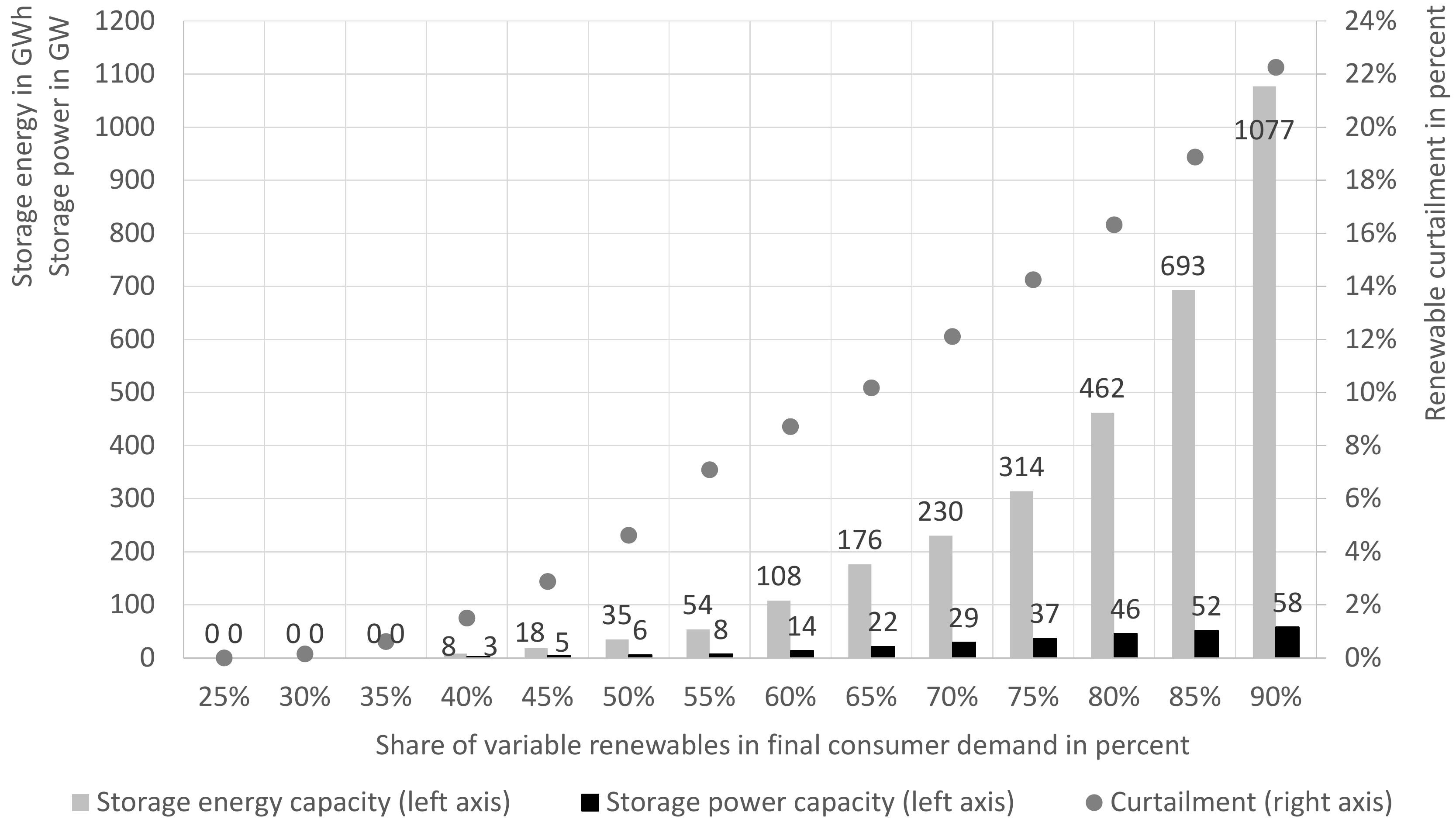}
\caption{Cost-optimal storage capacities and renewable curtailment rise with the minimum share of variable renewables. Yet they remain moderate.}
\label{fig:9_results_costmin}
\end{figure}

Figure~\ref{fig:9_results_costmin} summarizes the results of the parsimonious optimization model. Optimal storage capacities, both with respect to energy and power, rise with the share of variable renewable energy. However, overall storage requirements remain moderate. For instance, a storage energy capacity of~$35$~GWh suffices to achieve a share of~$50\%$~renewables in final demand. This is about two orders of magnitude less than in Sinn’s analysis, but slightly more than when targeting minimum storage requirements because the model considers additional values of storage. Yet it is still less than installed in Germany by 2018. Analogous findings prevail for other renewables shares.\footnote{Figure~\ref{fig:app_storage_base} in \ref{app:cost_optimal_base_years} provides results for alternative base years.} 

\vspace*{5pt}

When increasing the targeted renewables share, the optimal storage energy capacity grows much faster than the optimal storage power capacity. For~$50\%$~renewables, $35$~GWh energy are accompanied by about~$6$~GW storage power. Dividing energy by power yields an energy-to-power (E/P) ratio of about~$6$~hours. The E/P ratio is an important metric to characterize a storage technology and reflects its temporal layout: a $6$~hours storage is a typical short-to-medium-term storage to compensate diurnal fluctuations, such as of solar PV generation. If it is completely charged, it can generate electricity for~$6$~hours at maximum power rating.\footnote{The German pumped-hydro storage fleet has an E/P ratio of about~$7$~hours.} For higher renewables shares, the E/P ratio increases to reach about~$19$~hours for~$90\%$ renewables. This highlights the importance of considering both rather inexpensive storage energy and rather expensive storage power separately. Moreover, no need for a true long-term storage arises, that is, storing energy for weeks or months. 

Optimal endogenous renewable curtailment also grows with higher minimum renewables shares. As such, the economics of renewable electricity provide no reason why curtailment should be avoided. It can be more efficient not to use available renewable energy at times despite costly investment into wind and PV plants. The optimal solution combines conventional plants, storage, and renewables, part of which being curtailed at times.


\subsection{Extension: flexible sector coupling (\textit{power-to-x})}\label{subsec:costmin_p2x}
In future low-carbon energy systems, renewable electricity supply considered as surplus energy in the above framework is likely to be highly valuable for new uses. Merging electricity, heating, and transport sectors can not only provide flexibility for integrating variable renewables into the power market, but can also contribute to decarbonizing these other sectors \citep[]{Mathiesen.2015}. This concept, often referred to as \textit{sector coupling}, comprises using renewable electricity, for example, for residential heating \citep[see][for an overview of the recent literature]{Bloess.2018} or for electric mobility \citep{Richardson.2013}. Moreover, renewable electricity can also be used to produce other energy carriers, such as hydrogen or synthetic gaseous or liquid fuels, by means of electrolysis \citep{Schiebahn.2015}. Such sector coupling options are often referred to as \textit{power-to-x} (or P2X). They can lower electrical storage requirements if the new loads are sufficiently flexible and the additional demand can be shifted to periods in which renewable availability is high.

To illustrate this avenue in our model, we add a stylized additional electricity demand of a generic \textit{power-to-x} technology with a capacity of $n^x=50$~GW and $2,000$~full-load hours to our model, which corresponds to an additional annual energy demand of~$d^x=100$~TWh.\footnote{Full-load hours are an indicator for the annual use of the technology in terms of hours with demand at full capacity; for instance,~$2,000$~full-load hours are equal to~$4,000$~hours of use at half capacity.} For instance, this could be a fleet of electric vehicles: assuming a yearly electricity demand of $2,000$~kWh per vehicle, this would correspond to $50$~million vehicles. Alternatively, the additional demand may come from flexible electric heaters, or from electrolyzers converting renewable surplus electricity into hydrogen. For simplicity, we do not further restrict the timing of this additional consumption; it is only limited by the installed power capacity, i.e.~the \textit{power-to-x} demand is assumed to be highly flexible. In line with the literature, we generally assume that it is less costly to store \textit{x}, for instance heat or synthetic fuels, than electrical energy, rendering the timing of the \textit{power-to-x} generation more flexible. Moreover, we require the additional demand to be satisfied entirely by additional renewables. To this end, we augment Equation~\eqref{eq:3_distrib} to
\begin{align}
\gamma_hN^{r} = G^{r}_h + \overrightarrow{S}_h + C^{r}_h + \overrightarrow{X}_h \qquad \forall~h \tag{3b'} \label{eq:3'_distrib_p2x}
\end{align}

where the variable $\overrightarrow{X}_h$ is the hourly \textit{power-to-x} demand that is optimized endogenously. Two further equations restrict the hourly demand by the installed capacity~$n^x$~\eqref{eq:p2x_maxpower} and require that annual \textit{power-to-x} demand~$d^{x}$ is satisfied over the year~\eqref{eq:p2x_demand}. Otherwise, all model assumptions and equations are identical to Section~\ref{subsec:costmin_model}.
\begin{subequations}
\begin{align}
\overrightarrow{X}_h &\leq n^x \qquad \forall~h \label{eq:p2x_maxpower} \\
\sum_{h}\overrightarrow{X}_h &= d^{x} \label{eq:p2x_demand}
\end{align} 
\end{subequations}

\begin{figure}[htbp]
\centering
\includegraphics[width=1\linewidth]{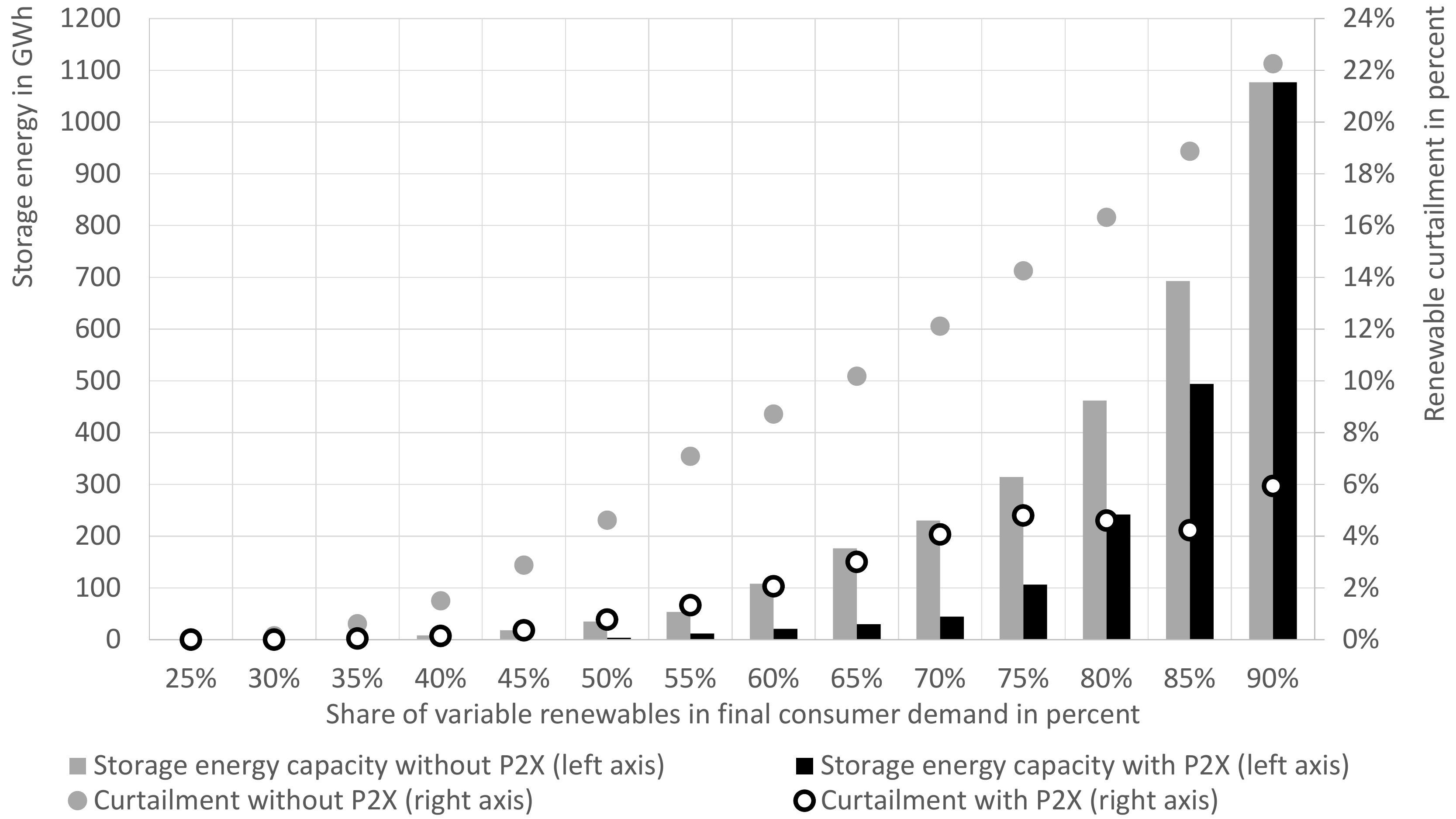}
\caption{Cost-optimal electrical storage capacities and renewable curtailment with an additional annual \textit{power-to-x} electricity demand of~$100$~GWh are substantially lower.}
\label{fig:10_results_costmin_p2x}
\end{figure}

Figure~\ref{fig:10_results_costmin_p2x} summarizes the results. For most renewables shares, both optimal storage capacities and renewable curtailment rates are substantially lower in case \textit{power-to-x} technologies are included. For instance, for~$50\%$~renewables, the storage energy capacity drops from~$35$~GWh to~$4$~GWh and renewable curtailment from~$5\%$ to below $1\%$. Electrical storage requirements are lower up to a renewables share of~$85\%$. In the~$90\%$ renewable case, storage needs are the same as in the case without \textit{power-to-x}, as the additional demand can be completely satisfied from renewable surplus generation that would otherwise be curtailed.

\vspace*{5pt}

The rationale is the following: in the~$50\%$~renewables case, wind and solar capacities rise from~$223$~GW to~$298$~GW to supply part of the additional \textit{power-to-x} demand. Another part of the additional demand is satisfied by renewable electricity previously curtailed or stored. Accordingly, both storage needs and renewable curtailment rates are substantially lower in most cases.\footnote{To be transparent, results strongly depend on the assumptions on capacities and full-load hours of the generic \textit{power-to-x} technology. See Figure~\ref{fig:app_p2x} in \ref{app:p2x_sensitivity}. The effect of \textit{power-to-x} on storage needs is zero or negative in this setting up to ~$4,000$~full load hours, which is a considerably high value. With the parameterization used here, the largest effect on electrical storage emerges between $1,500$ and $2,000$ full-load hours.}

As we assume \textit{power-to-x} to be perfectly flexible, the diminishing effects on storage requirements and renewable curtailment constitute an upper bound for less flexible real-world applications. However, they illustrate that flexible sector coupling could substantially drive down electrical storage needs. If additional flexible electricity demand emerges that contributes to decarbonizing other energy sectors, then renewable surplus generation becomes a valuable resource.\footnote{Likewise, more variable generation and, accordingly, wholesale market prices, may incentivize also the current electricity demand to become more temporally flexible in the long-run. This would have an analogous mitigative effect on storage needs.}


\section{Discussion}\label{sec:discussion}
Departing from \citet{Sinn.2017}, we implement small but relevant changes to move his the setup away from corner solutions. The impact is substantial and reduces storage needs up to two orders of magnitude. At the same time, our analysis remains stylized and tractable. In the following, we highlight further important points that researchers should consider when analyzing storage needs to integrate variable renewable electricity, both against the background of Sinn's analysis and the large body of academic literature \citep{brown.2018}. 

First, the definition of efficiency should be clarified when it comes to variable renewable energy sources. \citet{Sinn.2017} seems to refer to inefficiency as both curtailment of renewables and storage use as such; the first motivated by avoiding waste, the second by avoiding backup capacities (referred to as ``double structure buffering'').\footnote{See, for instance, Table~1 or Figure~8 in~\citet{Sinn.2017}. Moreover, the notion of ``wasting energy'' from renewables is questionable. On that note, one may also consider electricity not produced by conventional plants, that is, full-load hours smaller than $8,760$, as ``waste''. In both cases, marginal costs are zero.} However, a welfare economic approach should rather target the least-cost provision of electricity for given minimum renewable energy constraints. The result is a combination of conventional and renewable plants, storage, and curtailment of a certain amount of renewable energy. Why the one or the other should be ``inefficient'' is unclear. In the optimum, the marginal cost of further expanding storage, renewables that are curtailed at times, and conventional capacities is equal. Which shares of renewable energy are optimal when also considering external costs, for instance, arising from climate change, local emissions or land use change, whether the market achieves this solution, or which regulatory measures are required is another issue left for analysis and discussion elsewhere.

Second, electrical storage has values beyond arbitrage. It can provide firm capacity and thus reduce the need for backup plants (see Section~\ref{subsec:costmin_intuition}), provide balancing reserves and other ancillary services to maintain power system stability, and may also help mitigating grid constraints.

Third, other types of energy storage are relevant beyond pumped-hydro. These comprise batteries, which could become much cheaper in the future \citep{Schmidt.2017},
\footnote{For a sensitivity of our optimization model parameterized to lithium-ion batteries instead of pumped-hydro storage, see \ref{app:battery_sensitivity}. Results are qualitatively unchanged.
Batteries may gain additional relevance in the context of grid-integrated electric vehicles \citep{Budischak.2013}.} or power-to-gas storage. Specifically, different storage technologies have different costs for power and energy capacities. While batteries are relatively cheap in power, they are expensive in energy, and vice versa for power-to-gas. Thus, electrical storage technologies have different optimal E/P ratios: batteries are generally suited for short-term storage of a few hours, pumped hydro for around six to ten hours, and power-to-gas for longer periods. An optimal deployment of different storage types can address different types of renewable fluctuations, for example, intra-hourly, diurnal or even seasonal \citep{Safaei.2015,Scholz.2017,Zerrahn.2017}. Such a differentiated storage fleet also tends to be smaller and cheaper.

Fourth, scaling up historical feed-in time series of a fixed proportion of wind and solar power tends to over-estimate flexibility requirements. Both market forces and the renewables support scheme in Germany tend to incentivize renewable generation when prices are higher and supply is, accordingly, scarce. Such \textit{system-friendly} renewables comprise wind turbines that dis-proportionally produce electricity when wind speeds are low, both due to their location and technical layout \citep{May.2017}. A similar argument holds for solar PV panels, which may be directed such that they generate more electricity in morning or afternoon hours. Even more relevant, offshore wind turbines have much smoother generation patterns and higher full-load hours than onshore wind parks.\footnote{For instance, in Germany in 2016, offshore wind power had more than~$3,200$ full-load hours and a coefficient of variation of~$0.73$; for onshore wind power (solar PV), full-load hours were somewhat below~$1,600$ ($900$) and the coefficient of variation was~$0.83$ ($1.53$) \citepalias{OPSD.2017}.} The sensitivity toward the base year in Section~\ref{sec:replication} further illustrates the relevance of appropriate input data choices. Ideally, analyses should be based on bottom-up weather data covering as many years as possible \citep{Staffell.2016}.

Fifth, further electricity market integration across national borders generally yields smoother residual load. When balanced over greater geographical areas, the variability of wind, solar, and demand tends to be evened out \citep{Cebulla.2017,Fuersch.2013, Haller.2012,MacDonald.2016}. This results in smoother residual load patterns and lower storage requirements.

Finally, a temporally more flexible demand can also substitute electrical storage \citep{Denholm.2011, Pape.2014, Schill.2018}. If a more variable electricity supply triggers more volatile prices, and these prices are passed through to consumers, then the demand side should have increasing incentives to consume more flexibly and profit from arbitrage gains.


\section{Conclusions}\label{sec:conclusion}
The use of renewable energy is a major strategy to mitigate greenhouse gas emissions, reduce fossil fuel imports, and create a sustainable energy system. However, integrating growing shares of variable wind and solar power in electricity markets poses increasing challenges. Electrical storage is an important---albeit not the only---option to address the mismatching time profiles of variable renewable supply and electric demand. In a recent analysis, \citet{Sinn.2017} calculates storage needs in a German setting and finds vastly growing electrical storage requirements, already for renewable supply shares only moderately greater than currently the case in Germany. Based on these findings, he suggests that electrical storage may limit the further expansion of variable renewable energy sources.

While Sinn's illustrations deserve merit, the findings are not backed up by the literature. A large body of techno-economic studies conclude on substantially lower storage needs, also for high shares of variable renewables. An important reason for Sinn's deviating findings is that he only considers corner solutions---either no storage, resulting in vast renewable curtailment, or no curtailment, resulting in excessive storage requirements. 

We show that addressing these implicit assumptions matters: both results and conclusions change substantially. Our analysis, based on open-source tools and open data, concludes that storage needs are lower by up to two orders of magnitude. Our findings are in line with most of the literature. Cost-efficient solutions optimally combine renewable capacity expansion, renewable curtailment, and electrical storage. We also illustrate that electrical storage needs may decrease further if the electricity sector is broadened to also include flexible additional demand, for example related to heating, mobility or hydrogen production. While we demonstrate that such \textit{power-to-x} options may substantially change the picture, further and more detailed research on this avenue would be desirable.

All things considered, we conclude that electrical storage requirements do not limit the further expansion of variable renewable energy sources.


\section*{Acknowledgments}
We thank two anonymous reviewers for valuable remarks and Friedrich Kunz for extensive discussions. We also thank Christian von Hirschhausen, Julia Rechlitz, J\"orn Richstein, Fabian St\"ockl as well as the participants of the Mannheim Energy Conference 2018, the \"OGOR-IHS Workshop 2018 in Vienna and the BB$^{2}$ Bavarian Berlin Energy Research Workshop 2018 for many valuable comments. We further thank Mona Setje-Eilers for research assistance. The analysis was carried out in the context of the research project Kopernikus P2X, funded by the German Federal Ministry of Education and Research (grant number FKZ 03SFK2B1). Declarations of interest: none.


\pagebreak
\section*{References}
\bibliographystyle{plainnat}
\bibliography{References}


\pagebreak
\appendix
\setcounter{figure}{0}


\section{Literature review: storage power capacity requirements}\label{app:literature_power}
\begin{figure}[htb]
\centering
\includegraphics[width=1\linewidth]{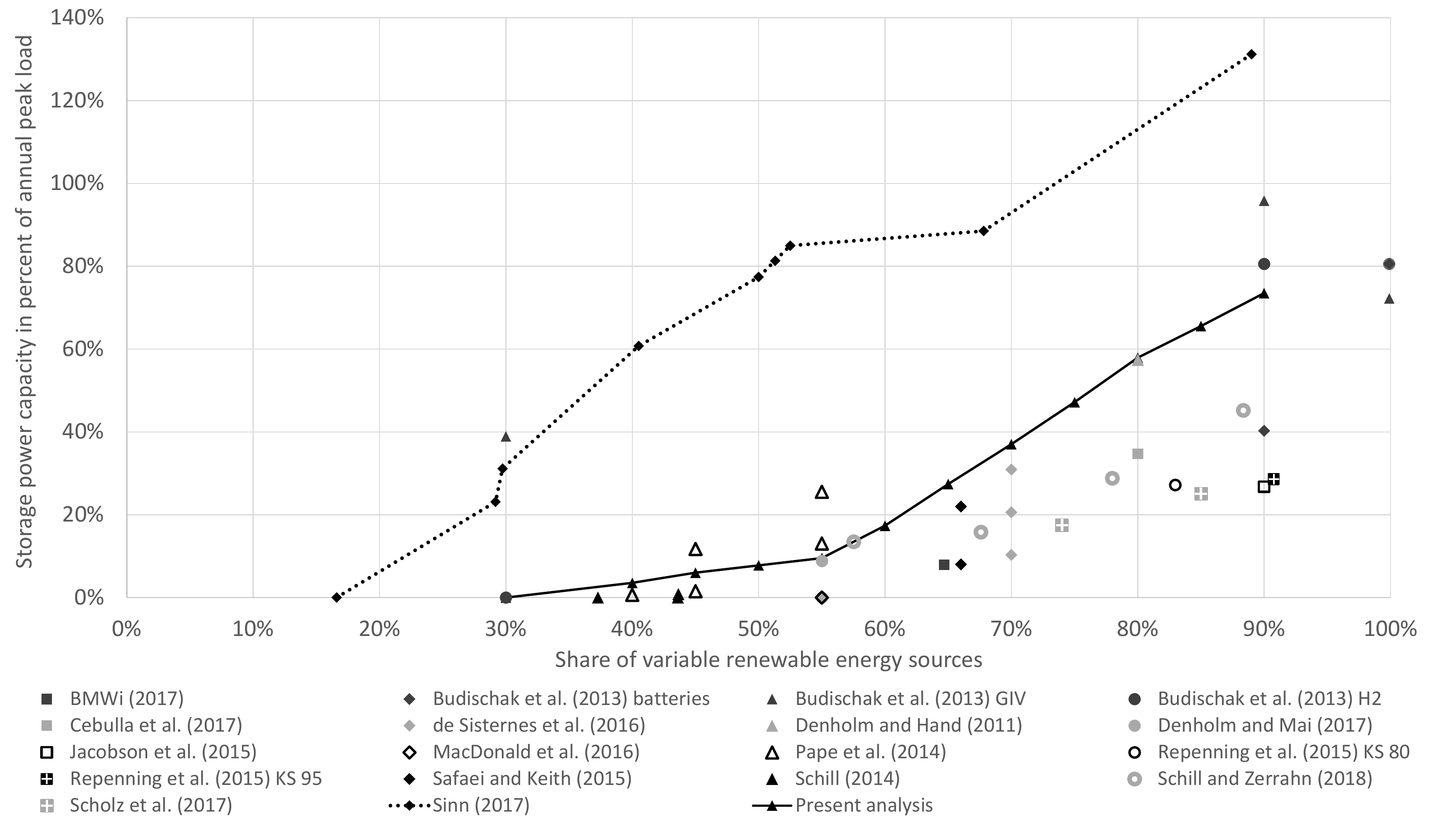}
\caption{The literature finds much lower storage power capacities than \citet{Sinn.2017}.}
\label{fig:app_survey_power}
\end{figure}

Figure~\ref{fig:app_survey_power} illustrates the comparison between storage power (i.e.~discharge) capacity requirements derived by \citet{Sinn.2017} and the literature. As \citet{Sinn.2017} does not explicitly mention any storage power capacities, we use the numbers from our replication of his calculations presented in Section \ref{sec:replication}. We also include our findings from the cost minimization approach. For comparability across studies, we normalize storage power capacities by dividing them by the system peak load. The system peak load arises in the hour with the highest demand. For Germany, it amounted to about~$79$~GW in 2014. As in the case of storage \textit{energy}, the literature finds much lower storage \textit{power} requirements than \citet{Sinn.2017}. 

\vspace*{5pt}

Figures \ref{fig:survey_energy_log} and \ref{fig:app_survey_power} require information not explicitly provided in several of the underlying studies. We calculate or infer missing data. Further, we select the most relevant cases. In the following, we provide additional information:

\begin{itemize}
	\item \citetalias{bmwi.2017}: Aggregated values for Germany and rest of Europe from \textit{Basisszenario}; peak load is not provided, but inferred from the ratio between peak load and yearly demand from \citet{Scholz.2017}.
	\item \citet{Budischak.2013}: Case with 2030 cost assumptions.
	\item \citet{Cebulla.2017}: Baseline; peak load is not provided, but inferred from the ratio between peak load and yearly demand from \citet{Scholz.2017}.
	\item \citet{Denholm.2017}: Minimum curtailment scenario, storage capacity of~$8.5$~GW and durations of 4, 8, and 12 hours.
	\item \citet{deSisternes.2016}: Scenarios without nuclear and with 10-hour storage; total yearly demand is not provided, but inferred from peak load using the ratio between peak load and yearly demand from \citet{Denholm.2011}.
	\item \citet{Jacobson.2015}: Yearly demand taken from Table 2 (without storage losses); storage including concentrating solar power. Only excess solar heat is shed; all surplus electricity is used for hydrogen generation.
	\item \citet{MacDonald.2016}: Scenario with low-cost renewables and high-cost natural gas; total yearly demand is not provided, but inferred from peak load using the ratio between peak load and yearly demand from \citet{Budischak.2013}.
	\item \citet{Pape.2014}: Only 2050 cases; variable renewable energy shares based on own estimations; peak load is not provided, but inferred from the ratio between peak load and yearly demand from \citet{Scholz.2017}.
	\item \citet{Repenning.2015}: Exogenous assumptions on pumped hydro storage capacity in Germany and Norway, the latter directly coupled via HVDC lines; no information on storage energy provided, we assume an E/P ratio of 8 hours.
	\item \citet{Safaei.2015}: Derived from Table S3, case with no dispatchable zero-emission generation; total yearly demand is not provided, but inferred from peak load using the ratio between peak load and yearly demand from \citet{Denholm.2011}.
	\item \citet{Schill.2014}: 2032 case without thermal must-run.
	\item \citet{Schill.2018}: Results for baseline assumptions.
	\item \citet{Sinn.2017}: Peak load and yearly demand as in present analysis.
	\item \citet{Scholz.2017}: Scenario with medium carbon price and equal shares of solar and wind.
\end{itemize}


\pagebreak
\setcounter{figure}{0}
\section{Sensitivity: optimal storage for different base years}\label{app:cost_optimal_base_years}
\begin{figure}[htbp]
\centering
\includegraphics[width=1\linewidth]{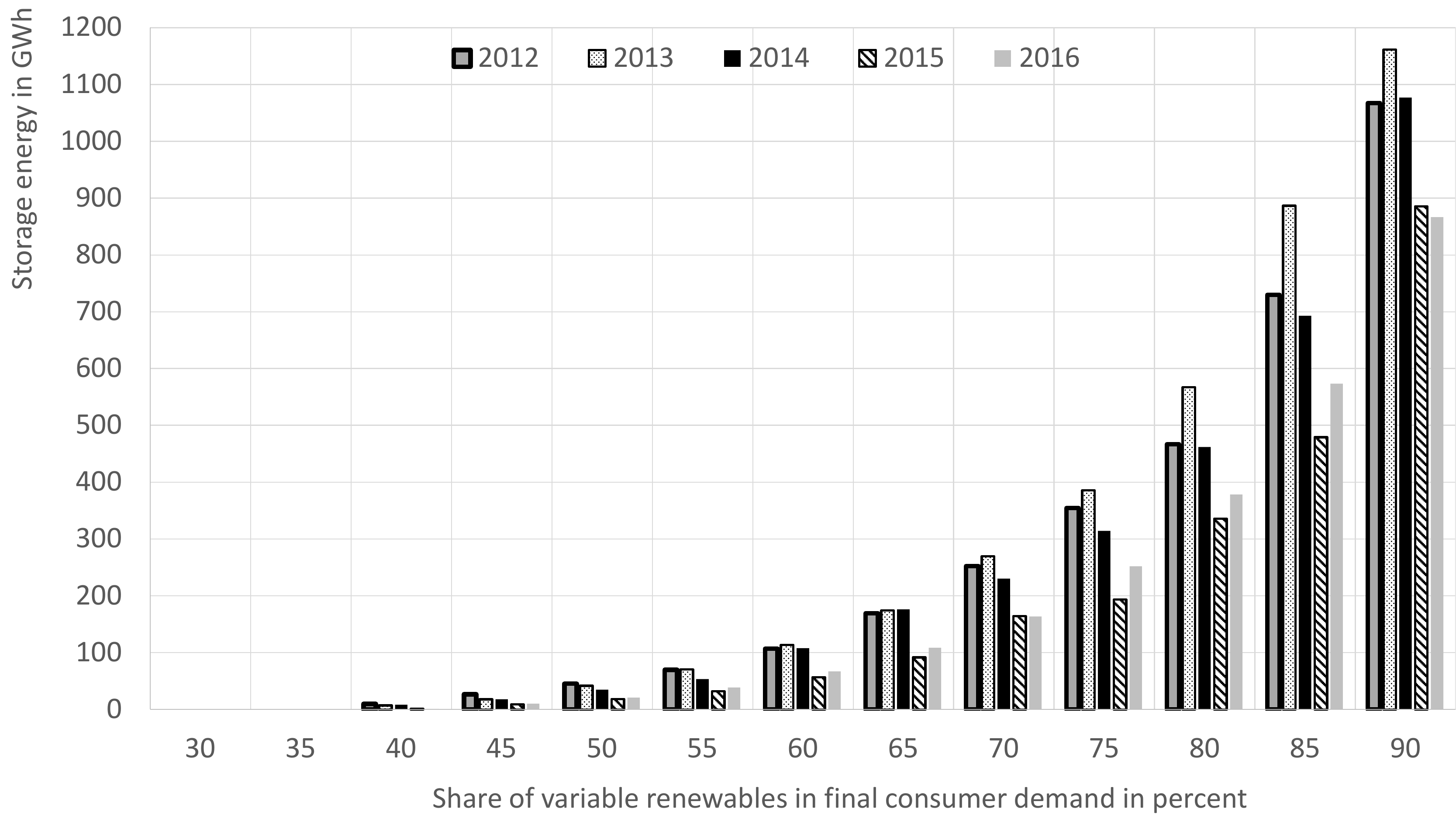}
\caption{Cost-minimal storage capacities are highly sensitive toward the choice of the base year. Results also substantially differ from the calculations that do not take costs into account (compare Figure~\ref{fig:2_baseyears}).}
\label{fig:app_storage_base}
\end{figure}

Figure~\ref{fig:app_storage_base} shows the cost-minimal storage energy capacities when using different base years. Base years 2015 and 2016 deliver much lower optimal storage capacities than the other base years. Smoother patterns of residual load are an important factor driving this result. 

\vspace*{5pt}


\pagebreak
\setcounter{figure}{0}
\section{Storage requirements in the optimization model: sensitivity with lithium-ion batteries}\label{app:battery_sensitivity}
\begin{figure}[htbp]
\centering
\includegraphics[width=1\linewidth]{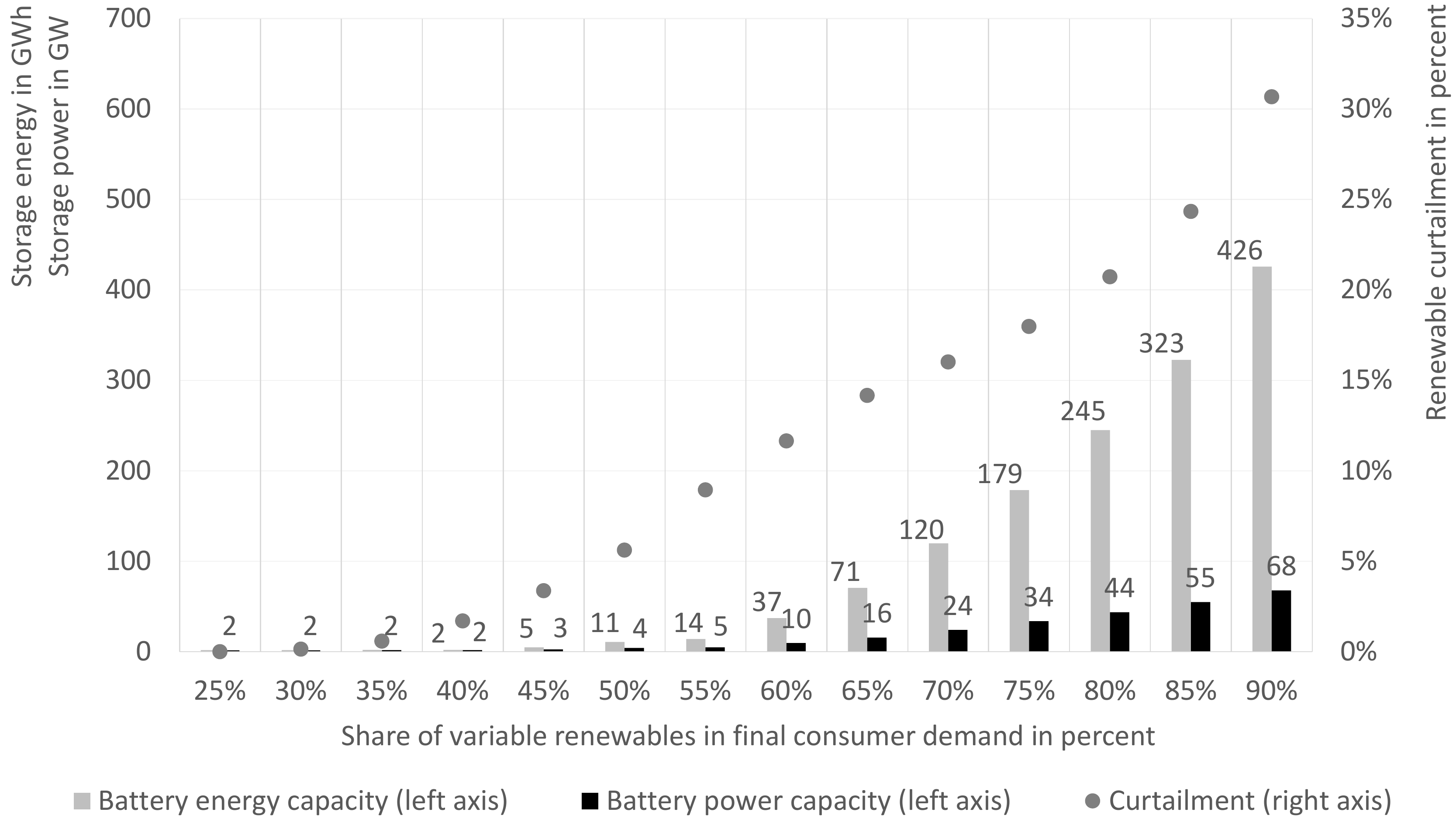}
\caption{Parameterizing storage to lithium-ion batteries in the numerical model leaves qualitative results unchanged.}
\label{fig:app_battery}
\end{figure}

Figure~\ref{fig:app_battery} plots cost-optimal storage energy and power capacities as well as curtailment rates for different shares of renewables in final demand for a sensitivity where the storage technology is parameterized to lithium-ion batteries instead of pumped hydro. Here, costs for storage power capacity are lower and costs for storage energy capacity are higher compared to the other calculations presented in this paper. In addition, the round-trip efficiency of lithium-ion batteries is higher than those of pumped-hydro storage.

\vspace*{5pt}

Three findings emerge: first, qualitative results are unchanged. Also here, storage energy requirements are substantially lower than in \citet{Sinn.2017}. Second, absolute storage energy deployment is lower compared to the case with pumped-hydro storage for most renewable shares, driven by higher specific costs. For instance, for $50\%$ variable renewables, they amount to $11$~GWh energy and $4$~GW power, opposed to $35$~GWh and $6$~GW in the pumped hydro case.
In contrast, the optimal renewable curtailment rate is higher: for instance, for $50\%$ renewables, it amounts to $6\%$, opposed to $5\%$; and for $80\%$ renewables, it amounts to $21\%$, opposed to $16\%$. Third, the total costs of providing electricity are also slightly higher, as the lower-cost option pumped-hydro storage is not available. 

Accordingly, pumped-hydro storage, which is often considered a medium-term storage technology ($5$--$10$hours), appears to be an appropriate technology choice in the context of this stylized analysis. However, as discussed in Section~\ref{sec:discussion}, an optimal portfolio would combine different types of electricity storage and further flexibility options to temporally align supply and demand in a cost-minimal way. To keep the exposure lean and focused, we abstain from devising a full-fledged analysis. \citet{Zerrahn.2017} provide a literature overview.


\pagebreak
\setcounter{figure}{0}
\section{Power-to-x: sensitivity with respect to different configurations}\label{app:p2x_sensitivity}
\begin{figure}[htbp]
\centering
\includegraphics[width=1\linewidth]{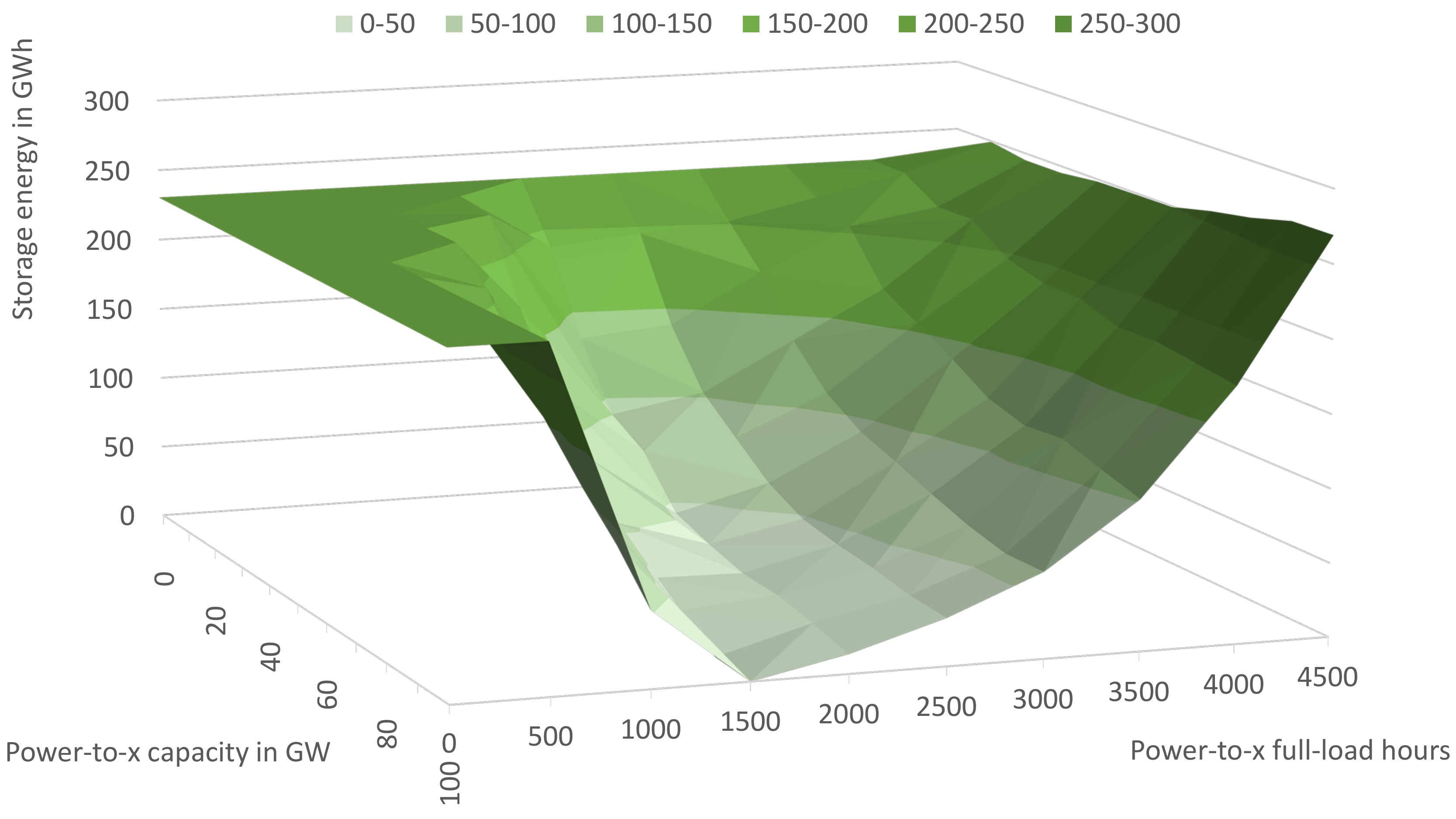}
\caption{Storage requirements for different \textit{power-to-x} settings for~$70\%$ variable renewables. The impact of additional flexible power-to-x demand on electrical storage requirements is largest between $1500$ and $2000$ full-load hours.}
\label{fig:app_p2x}
\end{figure}

Figure~\ref{fig:app_p2x} plots optimal storage energy capacities for different capacities and full-load hours of the generic \textit{power-to-x} technology. Specifically, medium full-load hours between around~$1,000$ and~$3,500$ can trigger substantially lower storage needs. For lower full-load hours, \textit{power-to-x} demand can largely be satisfied from renewable surpluses otherwise curtailed, so there is little or no effect on optimal storage capacity. For very high full-load hours, storage needs increase again. This is because of the increasing mismatch of the time profiles of additional \textit{power-to-x} demand and renewable availability, which triggers dis-proportional renewable capacity expansion, and in turn increases renewable  curtailment and the optimal amount of electrical storage.

\vspace*{5pt}


\end{document}